\documentclass[preprint,showpacs,preprintnumbers,amsmath,amssymb]{revtex4}

\usepackage{graphicx}
\usepackage{dcolumn}
\usepackage{bm}
\usepackage{textcomp}
\newcommand {\apgt} {\ {\raise-.5ex\hbox{$\buildrel>\over\sim$}}\ }
\newcommand {\aplt} {\ {\raise-.5ex\hbox{$\buildrel<\over\sim$}}\ }

\begin{document}
 
\title{Ion Crystal Metamorphoses in a Paul trap} 

\author{V. Ursekar, J. M. Silvester, Y. S. Nam, and R. Bl\"umel}  
\affiliation{Department of Physics, Wesleyan University, 
Middletown, Connecticut 06459-0155, USA}
  
\date{\today}

\begin{abstract} 
The standard second-order pseudo-oscillator potential
used in many analytical investigations of the properties
of ions stored in a Paul trap has serious limitations.
In this paper we show that ion-crystal configurations
exhibited by 2, 3, and 4 simultaneously stored ions in
a Paul trap are not predicted by the standard pseudo-oscillator
potential, but are all captured qualitatively
and quantitatively by an extended pseudopotential
derived in this paper.
The power of our extended pseudopotential extends in particular
to the prediction of the border lines between different
crystal configurations (morphologies) in the Paul trap's
$a$, $q$ stability diagram.
In the three- and four-ion cases, several of the ion-crystal structures
predicted by our improved pseudopotential have never been
observed experimentally before.
We present them here as a challenge for experiments.
\end{abstract}


           

\maketitle
\section{Introduction}
\label{INTRO} 
Periodically driven systems abound in 
atomic, molecular, and optical (AMO) physics. 
Examples are laser interactions with atoms 
and ions 
\cite{AE,SSL}, 
Rydberg atoms in strong external 
microwave fields 
\cite{BK,KL,RB1,RB2}, 
excitation of molecular rotation by periodic 
microwave pulses 
\cite{BFS}, 
and charged 
particles in electrodynamic traps 
\cite{Paul1,PKG,Nature}. 
Treating the radiation field classically, 
all these periodically driven AMO systems 
are governed by a time-periodic Hamiltonian 
\begin{equation}
\hat H(t+T) = \hat H(t),
\label{INTRO1}
\end{equation}
where $T$ is the period of the driving field. 
The wavefunction of the system then satisfies 
the periodically driven Schr\"odinger equation 
\begin{equation}
i\hbar \frac{\partial |\psi(t)\rangle}{\partial t} = 
\hat H(t) |\psi(t)\rangle, 
\label{INTRO2}
\end{equation}
where $\hbar$ is Planck's constant. This equation 
may be solved formally by the linear, unitary 
time evolution operator 
\begin{equation}
\hat U(t') |\psi(t)\rangle = |\psi(t'+t)\rangle.  
\label{INTRO3}
\end{equation}
Because of the linearity of (\ref{INTRO2}), the 
time-evolution operator $\hat U$ satisfies the 
operator identity
\begin{equation}
\hat U(t') \hat U(t) = \hat U(t'+t). 
\label{INTRO3a}
\end{equation}
Since $\hat U(t)$ is unitary for all $t$, 
$\hat U(T)$ is unitary and may be written as 
\begin{equation}
\hat U(T) = \exp\left(\frac{i}{\hbar}\hat W T\right),
\label{INTRO4}
\end{equation}
where $\hat W$ is a Hermitian operator, 
also known as the 
{\it quasi-energy} operator \cite{QE}. 
With the help of $\hat W$ we define the propagator 
\begin{equation}
\hat P(t) = \exp\left(\frac{i}{\hbar}\hat W t\right),
\label{INTRO4a}
\end{equation}
which interpolates smoothly between the values of 
$\hat U(NT)$, $N$ integer, and agrees with $\hat U(t)$
at integer multiples of the driving period $T$. 
The 
quasi-energy operator $\hat W$, with its 
associated propagator $\hat P(t)$, is all we need to 
obtain a stroboscopic description of the 
time evolution of the wavefunction at 
multiples of the driving period $T$: 
\begin{equation}
|\psi(NT)\rangle = 
\hat U(NT)|\psi(0)\rangle = 
\exp\left(\frac{i}{\hbar}\hat W NT\right) |\psi(0)\rangle = 
\hat P(NT) |\psi(0)\rangle. 
\label{INTRO5}
\end{equation}
We may use the operator $\hat W$ to obtain 
a smoothed time evolution 
of $|\psi\rangle$ on the macro-scale $T$ according to 
\begin{equation}
|\psi(t)\rangle \approx |\varphi(t)\rangle = 
\hat P(t) |\psi(0)\rangle, 
\label{INTRO6}
\end{equation}
where, in line with our stroboscopic picture, 
the smoothed motion described by $|\varphi(t)\rangle$ agrees 
with the exact time evolution of $|\psi(t)\rangle$ at 
multiples of the drive-period $T$ according to 
\begin{equation}
|\varphi(NT)\rangle =
|\psi(NT)\rangle . 
\label{INTRO7}
\end{equation}
Because $|\varphi(t)\rangle$ describes the motion on the 
macro-scale $T$, we call the motion represented by 
$\hat P(t)$ and its associated wavefunction 
$|\varphi(t)\rangle$ the {\it macromotion}. In order 
to obtain the exact time evolution, but keeping the 
convenient description provided by $|\varphi(t)\rangle$ 
on the macro-scale, we define the reduced propagator 
\begin{equation}
\hat Q(t) = \exp\left[-\frac{i}{\hbar}\hat W (t\,{\rm mod}\,T)\right]
\hat U(t\,{\rm mod}\, T), 
\label{INTRO8}
\end{equation}
which is defined on the time interval $[0,T]$. 
With the help of $\hat W$ and $\hat Q(t)$, the 
complete time-evolution operator $\hat U(t)$ may now 
be factored exactly into two parts, 
\begin{equation}
\hat U(t) =  \hat P(t)\, \hat Q(t), 
\label{INTRO9}
\end{equation}
where the first factor, $\hat P(t)$, 
describes the macromotion and the second factor, 
$\hat Q(t)$,  
describes the motion on the fine scale between multiples 
of $T$. Therefore, we call the motion described by $\hat Q(t)$ 
the {\it micromotion}. Defining $t=NT+\tau$, where 
$0\leq \tau < T$, it can be shown immediately that 
(\ref{INTRO9}) is indeed exact: Because of the 
definition of $\tau$, we have $\tau=t \bmod T$, 
and therefore 
\begin{align}
\hat P(t)\hat Q(t) &=  
\exp\left(\frac{i}{\hbar}\hat W (NT+\tau)\right) 
\exp\left(-\frac{i}{\hbar}\hat W \tau\right) \hat U(\tau)
\nonumber\\ 
&= \exp\left(\frac{i}{\hbar}\hat W NT\right) \hat U(\tau) 
\nonumber \\ 
&= \hat U(NT)\hat U(\tau) = \hat U(NT+\tau) = 
\hat U(t), 
\label{INTRO10}
\end{align}
where we used the operator identity (\ref{INTRO3a}). 
The separation into a slow macromotion, described by 
the quasi-energy operator $\hat W$ and its associated 
propagator $\hat P$, and a fast micromotion, 
described by the propagator $\hat Q$, is extremely useful 
in all cases in which we are more interested in the large-scale 
time evolution of a driven system than in its precise, 
in most cases extremely high-frequency 
behavior between driving periods. Therefore, 
looking at the system on the coarse-grained time scale $T$, 
governed by the interpolating time-evolution operator $\hat P$, 
we have succeeded in transforming 
an {\it explicitly time-dependent} system, 
described by the Hamiltonian $\hat H(t)$, containing 
time-periodic drive terms, 
into a {\it time-independent} system,
described by the new quasi-energy Hamiltonian $\hat W$. 
The analogy between $\hat W$ and a conventional time-independent 
Hamiltonian is further strengthened by the observation that 
in many cases, at least to lowest order, $\hat W$ may 
be written as the sum of a kinetic-energy operator $\hat K$ 
and a potential-energy operator, called the {\it pseudopotential}, 
or, in the AMO context, the ponderomotive potential \cite{EMP}. 

Formally, and in a physics context on the quantum level, 
this separation expresses the mathematical content of 
{\it Floquet's Theorem} \cite{Floquet,AS}, a 19th century 
theorem, first derived in connection with periodically 
driven linear systems of equations. 
 
The usefulness of 
the separation (\ref{INTRO9}) of the time-evolution operator 
into slow macromotion and fast micromotion components 
cannot be overstated. It is the basis of much of 
quantum chaos research \cite{Stock} with its implications for 
AMO physics \cite{QCR,AMOQE}, where the focus is on the spectral 
properties of $\hat W$ \cite{SPECP}. 

For many driven AMO systems, in
particular for strong driving fields and large
quantum numbers, a classical description of the driven 
atomic system
makes sense \cite{BK,KL,RB1,RB2}.
Motivated by the corresponding quantum systems, we are 
led to the question: Is an exact quasi-energy description also possible
for these classical AMO systems? 
At first glance, the answer seems to be ``no'', 
since these classical
systems are, in general, non-linear, while Floquet's Theorem, 
and all the procedures discussed above, are applicable 
only for linear (quantum) systems. However, contrary to 
expectations, a classical separation into a macromotion part 
and a micromotion part is indeed possible. In particular, 
it is possible to construct a classical pseudopotential \cite{LL} 
that averages over the fast, classical time scales and 
results in an effective pseudo-Hamiltonian that is the 
classical analog of $\hat W$. However, these procedures 
are not exact. 
While, as shown above, it is straightforward on 
the quantum level to exactly and explicitly separate 
the time-evolution operator into a slow, macro-motion component, 
and a fast, micro-motion component, a similar, generally valid 
decomposition, in analogy to the quantum procedures, is not 
known on the classical level. Usually, in particular in 
connection with ion-trapping systems \cite{Paul1,PKG,Nature}, 
the first-order pseudopotential, as constructed in \cite{LL}, 
is used. This, however, as shown 
in \cite{2ION,HB1,MB}, may not 
capture qualitatively important features even on the 
slow, macro-scale. Therefore, using the trapping 
of up to four ions in a Paul trap as our example, 
the purpose of this paper is 
to outline a generalizable procedure that is capable of 
constructing higher-order pseudopotentials to better 
approximate the time-independent Paul-trap pseudo-Hamiltonian. 
We will show that our improved pseudopotential is capable 
of predicting crystalline structures (morphologies) 
of ions in the Paul trap that are 
completely beyond the predictive power of the simple, 
lowest-order pseudopotential constructed according to 
\cite{LL}. We consider the discovery of these unconventional 
three- and four-ion morphologies not only as an illustration 
of the predictive power of our improved pseudopotentials,
but also as a challenge for experimentalists to confirm 
the existence of these exotic structures in the lab. 

Our paper is structured as follows.
In Sec.~\ref{TDEM} we introduce the time-dependent equations of motion of 
ions simultaneously stored in a Paul trap. These equations of 
motion are the basis and starting point of all of our work 
described in this paper. In this section we also give a brief 
introduction into the pseudopotential method, illustrated for the case 
of two simultaneously stored ions. 
In Sec.~\ref{secIIa}, we develop a general mathematical procedure
that may be applied to a system of charged particles stored in the Paul trap.
In Sec.~\ref{secIIb}, and going beyond the 2-ion case 
discussed in \cite{2ION,HB1,MB}, 
we apply the formalism developed in Sec.~\ref{secIIa}
to the specific case of three stored ions
to obtain an improved pseudopotential for this case.
In Sec.~\ref{secIII} we revisit the case of two particles
in the Paul trap to highlight the power of the improved pseudopotential,
as this is the simplest possible case in the presence of an interparticle potential.
In particular, we will show how the improved pseudopotential reveals
new physics, beyond what is predicted by the standard pseudopotential.
Then, in Sec.~\ref{secIV}, together with the pseudopotential expression
derived in Sec.~\ref{secII}, and the demonstrated method of revealing
new physics in Sec.~\ref{secIII}, we investigate the three-particle case
in detail, where we report morphology boundary lines of different
ion-crystal configurations in the Paul trap.
In Sec.~\ref{secV} we present numerical results on the four-particle case
and address how our general formalism may be applied.
In Sec.~\ref{secVI} we discuss our results and in Sec.~\ref{secVII}
we summarize and conclude our paper.
 
 
\section{Time-dependent equations of motion}
\label{TDEM}
\begin{figure}
\centering
\includegraphics[scale=0.6,angle=0]{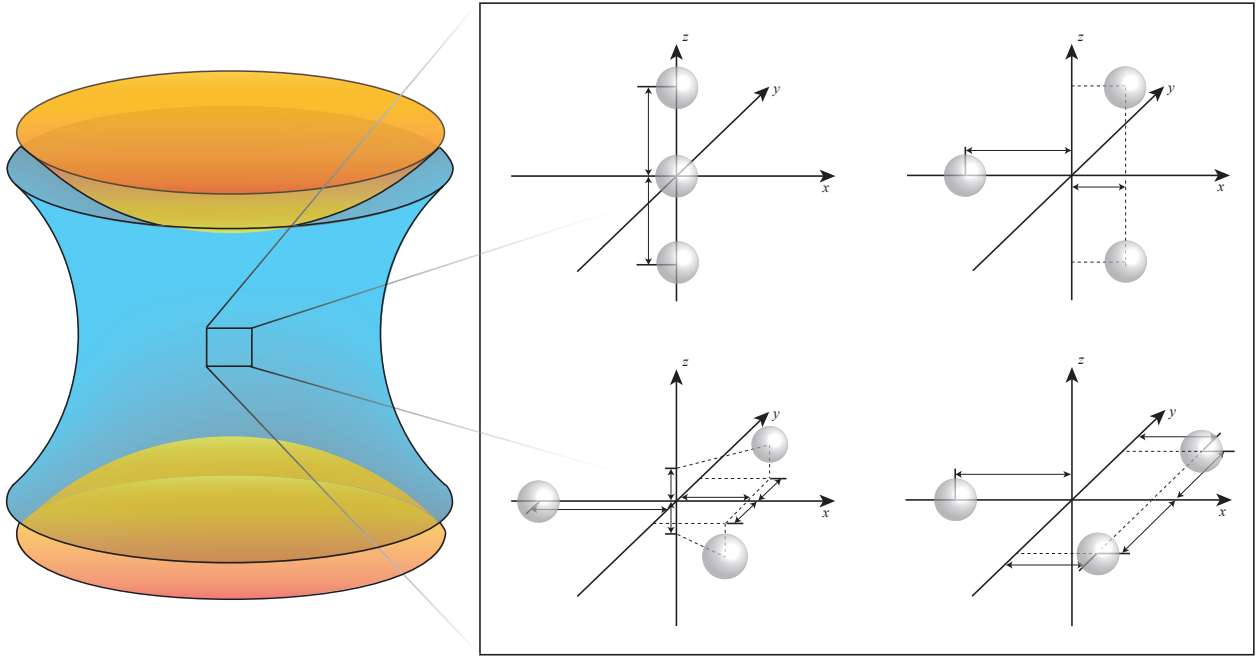}
\caption {\label{fig1}
(Color Online)  Electrode configuration of a Paul trap (left)
together with four possible orientations of 
a three-particle Coulomb crystal in a Paul trap (box on the right).
The Paul trap consists of two end-cap electrods (orange)
at top and bottom and a ring electrode (blue), 
both of which are connected to ac and dc voltage sources (not shown).
Shown to the right, inside the box, are the observed three-particle crystal 
configurations in a Paul trap,
obtained from numerical simulations.
The four configurations are rod (top-left),
pop-out (top-right), tilt (bottom-left), and
planar (bottom-right),
where, as indicated by their positions, illustrated
by arrows and dashed lines, the center of mass of
each of the configurations is located at the origin,
i.e., the center of the Paul trap.
}
\end{figure}

As shown in Fig.~\ref{fig1}, a Paul trap \cite{Paul1,PKG,Nature} 
consists of a concentric arrangement of 
three conducting surfaces, a hyperbolic 
ring electrode and two electrically connected 
hyperbolic end-cap electrodes. As described in the 
literature \cite{Paul1,PKG,Nature}, applying a 
combination of ac and dc voltages between the 
ring- and end-cap electrodes produces a time-dependent quadrupole 
potential capable of storing $N$ ions simultaneously 
for very long periods of time. 
Storage times ranging from hours \cite{BlWe}
to days \cite{WSL} with up to $10^5$ simultaneously
stored ions \cite{BlWe} have been reported. 
 
In this paper, we treat the stored ions in the 
classical approximation. Even if the ions are 
in their crystalline state \cite{Nature,MPQ1,NIST1}, 
this approximation is expected to be 
excellent due to the large distances between 
ions in conventional traps (of the order of $\mu$m 
\cite{Nature,MPQ1,NIST1}) and their relatively 
large temperatures (no effort is usually made 
to cool ion crystals to their quantum ground state, 
although this has been demonstrated \cite{Wineland1} 
to be possible in the single-ion case). 
 
Measuring time in units of 
\begin{equation}
t_0 = \frac{2}{\omega}, 
\label{TDEM1}
\end{equation}
where $\omega$ is the angular frequency 
of the driving ac voltage, 
and distances in units of 
\begin{equation}
l_0 = \left( \frac{Q^2}{\pi\epsilon_0 m \omega^2} \right)^{1/3}, 
\label{TDEM2}
\end{equation}
where $Q$ is the charge of the stored particles, 
$m$ is their mass, and $\epsilon_0$ is the 
permittivity of the vacuum, the classical equations of 
motion in an ideal Paul trap \cite{Paul1,PKG} are 
given by 
\begin{equation}
\ddot{\vec R}_i + \gamma \dot{\vec R}_i + 
[a+2q\cos(2 \tau)]
 \left( \begin{matrix} X_i \\ Y_i \\ -2 Z_i \end{matrix} \right) 
 =   \sum_{\substack{j=1\\j\neq i}}^N 
 \frac{\vec R_i - \vec R_j}{|\vec R_i - \vec R_j |^3}, \ \ \ i=1,\ldots,N,
\label{TDEM3}
\end{equation} 
where ${\vec R}_i=(X_i,Y_i,Z_i)$ is the position vector of 
ion number $i$ in units of $l_0$, $\gamma$ is a damping constant, 
$\tau$ is the time in units of $t_0$, and $a,q$, given by 
\begin{equation}
a = \frac{8Q V_{\text dc}}{m\omega^2 (r_0^2+2z_0^2)} ,
\ \ \ q=   \frac{4Q V_{\text ac}}{m\omega^2 (r_0^2+2z_0^2)} ,
\label{TDEM4}
\end{equation} 
are the dimensionless control parameters of the trap, 
proportional to the applied dc and ac voltages, $V_{\text dc}$ and 
$V_{\text ac}$, respectively, and  $r_0$ and $z_0$ 
in (\ref{TDEM4}) are the distances of the ring electrode and 
the end-cap electrodes from the center of the trap, 
respectively. 
 
 The damping constant $\gamma$ in (\ref{TDEM3}) deserves a few comments. 
It may be used two ways. (i) Frequently in the literature \cite{Nature}, 
this 
constant is used to simulate actual ion cooling present in 
ion-trapping experiments. (ii) Our use of this damping constant is 
different. We are interested in the properties of ion crystals in the ideal 
Paul trap in the absence of damping. Therefore, 
in all of our simulations, we use $\gamma$ only as an auxiliary 
device to cool ion clouds \cite{Nature,EBJ} 
into crystals. Once the 
crystals have formed, we slowly switch off $\gamma$ so that 
we may assess ion-crystal properties for $\gamma=0$. 
It is known that ion crystals, once in 
the crystalline state, and even though they are driven 
by the trap's ac field, do not show any energy 
gain \cite{Nature,Kappler}. 
Therefore,  
in the absence of external heat sources, and even for 
$\gamma=0$, they remain in their crystal state forever. 
In the following, wherever in this paper
we refer to simulations of 
the equations of motion (\ref{TDEM3}), we imply that 
the $\gamma$-switch-off method just described is used 
to produce ion crystals in our ideal Paul trap. 
Since $\gamma=0$ at the end of our ion-crystal formation process, 
and since ion crystals are stable in the absence of $\gamma$, 
none of the crystal properties reported in this paper, 
as it should be, depend on $\gamma$. 
 
The set of equations of motion (\ref{TDEM3}) is the 
basis of our work described in this paper. 
Whenever we need to compare the predictions of our analytical 
approximations to the predictions of the exact dynamics, we 
use numerical solutions of the equations of motion 
(\ref{TDEM3}), obtained by integrating (\ref{TDEM3}) 
with the help of standard 4th- and 5th-order 
Runge-Kutta differential-equation solvers \cite{NumRec}. 
 
A first analytical stab at the equations of motion 
(\ref{TDEM3}) was done long ago using the method of 
averaging \cite{LL} to derive a time-averaged, 
time-independent pseudopotential, the 
standard pseudopotential 
\begin{equation}
U_{\text \rm eff}^{(s)} = \frac{1}{2} 
\sum_{i=1}^N\left[ \mu_x^2 X_i^2+\mu_y^2 Y_i^2 
+ \mu_z^2 Z_i^2 \right] + 
  \frac{1}{2} \sum_{\substack{i,j=1\\j\neq i}}^N  
  \frac{1}{|{\vec R}_i-{\vec R}_j|} , 
\label{TDEM5}
\end{equation} 
in which the $N$ ions are confined, despite their 
repulsive Coulomb interactions, subjected to 
the standard pseudo-oscillator with oscillator 
frequencies \cite{Kappler}
\begin{equation}
\mu_x = \mu_y=  \left( a+\frac{1}{2} q^2\right)^{1/2},
\ \ \ \mu_z = [2(q^2-a)]^{1/2}. 
\label{TDEM6}
\end{equation} 
In some regions of the $(q,a)$ control-parameter space, this 
potential is excellent and describes crystal configurations 
very well. In other parts of the $(q,a)$ control-parameter space 
the standard pseudopotential (\ref{TDEM5}) 
fails completely, even 
missing, as we will see, most of the 
possible crystal morphologies predicted by (\ref{TDEM3}). 
For the three-ion case 
this is illustrated in the box of 
Fig.~\ref{fig1}, which shows the four possible morphologies 
of a three-ion crystal, predicted by (\ref{TDEM3}), of which 
one of them (the tilted crystal)
is completely missed by the standard pseudopotential 
(\ref{TDEM5}) (for more details see below). 
The inadequacy of the standard pseudopotential 
in this respect was first pointed out in 
\cite{2ION,HB1,MB} for the two-ion case, 
and an adequate, improved pseudopotential, capable of 
describing all two-ion morphologies, was derived 
in \cite{2ION,MB}. It is 
the purpose of this paper to improve the 
standard pseudopotential (\ref{TDEM5}) such that it is 
powerful enough to predict, in addition to 
the two-ion case, all crystal morphologies and 
transition boundaries between them in $(q,a)$ 
control-parameter space for three and four simultaneously 
stored particles. We will see that 
the exact equations of motion (\ref{TDEM3}) 
predict a rich menagerie of exotic three- and four-particle 
crystal morphologies that are all captured by our 
improved pseudopotentials derived in the following 
sections. Our methods are generalizable and, 
in addition to the Paul trap, applicable 
to many other periodically driven AMO systems. 


\section{Three-particle pseudopotential}
\label{secII}
Since the improved pseudopotential for the two-ion case is already
known \cite{2ION,MB}, we focus in this section on the derivation
of the improved three-particle pseudopotential.
Following this section, the formal machinery will be in place
to apply our results to the two-ion case (Sec.~\ref{secIII}),
the three-ion case (Sec.~\ref{secIV}), and the four-ion case (Sec.~\ref{secV}).

\subsection{Derivation of a generalized pseudopotential in a driven system}
\label{secIIa}
 
We start with the following set of three coupled differential equations
\begin{align}
\label{dynamic}
&m_X \ddot{X} = -U_X(X,Y,Z)-k_X X \cos(\omega t), \nonumber \\
&m_Y \ddot{Y} = -U_Y(X,Y,Z)-k_Y Y \cos(\omega t), \nonumber \\
&m_Z \ddot{Z} = -U_Z(X,Y,Z)-k_Z Z \cos(\omega t),
\end{align}
where $m_I$, $I \in \{X,Y,Z\}$, denote effective masses,
$U_I=\partial U/\partial I$, and $k_I$ are constants.
The system (\ref{dynamic}) is generic and describes 
a variety of dynamical systems, of which one is the Paul trap.
The treatment of a general, $n$-coordinate system
is available in Appendix~\ref{appA}.
Here, we investigate the case $n=3$, which will be of particular importance
in connection with the three-particle morphology to be discussed
in detail in Sec.~\ref{secIIb}.

Following \cite{LL}, assuming that
the $\cos(\omega t)$ terms in (\ref{dynamic}) oscillate fast
compared to the motion governed by $U$,
we may write
\begin{align}
\label{motion}
& X(t) = x(t) + \xi\cos(\omega t), \nonumber \\
& Y(t) = y(t) + \eta\cos(\omega t), \nonumber \\
& Z(t) = z(t) + \zeta\cos(\omega t),
\end{align}
where $x,y,z$ denote the slowly varying macromotion coordinates
and $\xi,\eta,\zeta$ denote the amplitudes of the rapidly oscillating micromotion
in their respective directions.
The micromotion amplitudes are assumed to vary slowly
over the time scale of the macromotion and hence
may be assumed to be constant over a cycle of the driving field.
Assuming now that the micromotion amplitudes are small,
we may expand $U(X,Y,Z)$ in the micromotion amplitudes 
up to second order 
according to
\begin{align}
\label{pot1}
U(X,Y,Z) \approx \,\, & U(x,y,z) + U_x\xi\cos(\omega t) + U_y\eta\cos(\omega t) 
                                                   + U_z\zeta\cos(\omega t) \nonumber \\
                  + & \frac{1}{2}U_{xx}\xi^2\cos^2(\omega t) 
                   + \frac{1}{2}U_{yy}\eta^2\cos^2(\omega t) 
                   + \frac{1}{2}U_{zz}\zeta^2\cos^2(\omega t) \nonumber \\
                 + & U_{xy} \xi\eta \cos^2(\omega t)
                   + U_{yz} \eta\zeta \cos^2(\omega t)
                   + U_{zx} \zeta\xi \cos^2(\omega t),
\end{align}
where, for $i,j \in \{x,y,z\}\, (I,J \in \{X,Y,Z\})$, $U_{i}$ or $U_{ij}$ 
denote, in their respective order, $\partial U/\partial I$ 
or $\partial^2 U/\partial I \partial J$,
evaluated at $X = x, Y= y$, and $Z = z$.
The first derivative of (\ref{pot1}) in $X$, evaluated at 
$(x,y,z)$, then reads
\begin{align}
\label{fdpot1}
U_X \approx \,\, & U_x + U_{xx}\xi\cos(\omega t) + U_{yx}\eta\cos(\omega t) 
               + U_{zx}\zeta\cos(\omega t) \nonumber \\
             + & \frac{1}{2} U_{xxx}\xi^2\cos^2(\omega t)
                + \frac{1}{2}U_{yyx}\eta^2\cos^2(\omega t) 
                + \frac{1}{2}U_{zzx}\zeta^2\cos^2(\omega t) \nonumber \\
             + & U_{xyx} \xi\eta \cos^2(\omega t)
                + U_{yzx} \eta\zeta \cos^2(\omega t)
                + U_{zxx} \zeta\xi \cos^2(\omega t)
\end{align}
and similar expressions may be 
obtained straightforwardly for $Y$ and $Z$.
Together with (\ref{motion}), inserting (\ref{fdpot1}) 
and its corresponding $Y$ and $Z$ parts into (\ref{dynamic}) 
and equating like powers of $\cos(\omega t)$, we obtain 
\begin{align}
&m_X \omega^2 \xi = U_{xx} \xi + U_{yx} \eta + U_{zx} \zeta + k_X x, \nonumber \\
&m_Y \omega^2 \eta = U_{yy} \eta + U_{zy} \zeta + U_{xy} \xi + k_Y y, \nonumber \\
&m_Z \omega^2 \zeta = U_{zz} \zeta + U_{xz} \xi + U_{yz} \eta + k_Z z. 
\end{align}
Solving for the micromotion amplitudes, we obtain
\begin{align}
\label{micro}
\xi = \{ &k_X x [(m_Y\omega^2-U_{yy})(m_Z\omega^2-U_{zz})-U_{yz}^2] \nonumber \\
         +&k_Y y [U_{yx}(m_Z\omega^2-U_{zz})+U_{xz}U_{yz}] \nonumber \\
         +&k_Z z [U_{zx}(m_Y\omega^2-U_{yy})+U_{xy}U_{zy}] \}/\Delta, \nonumber \\
\eta = \{ &k_Y y [(m_Z\omega^2-U_{zz})(m_X\omega^2-U_{xx})-U_{zx}^2] \nonumber \\
         +&k_Z z [U_{zy}(m_X\omega^2-U_{xx})+U_{yx}U_{zx}]  \nonumber \\
         +&k_X x [U_{xy}(m_Z\omega^2-U_{zz})+U_{yz}U_{xz}]\}/\Delta, \nonumber \\
\zeta = \{ &k_Z z [(m_X\omega^2-U_{xx})(m_Y\omega^2-U_{yy})-U_{xy}^2] \nonumber \\
         +&k_X x [U_{xz}(m_Y\omega^2-U_{yy})+U_{zy}U_{xy}]  \nonumber \\
         +&k_Y y [U_{yz}(m_X\omega^2-U_{xx})+U_{zx}U_{yx}]\}/\Delta,
\end{align}
where
\begin{align}
\label{microD}
\Delta = & (m_X\omega^2-U_{xx})(m_Y\omega^2-U_{yy})(m_Z\omega^2-U_{zz}) \nonumber \\
&-(m_X\omega^2-U_{xx})U_{yz}^2-(m_Y\omega^2-U_{yy})U_{zx}^2 \nonumber \\
&-(m_Z\omega^2-U_{zz})U_{xy}^2-2U_{xy}U_{yz}U_{zx}.
\end{align}
This time, together with (\ref{motion}) and (\ref{fdpot1}),
averaging (\ref{dynamic}) over one cycle of the driving field, results in
\begin{align}
m_X \ddot{x} = & -U_x -\frac{1}{4} U_{xxx} \xi^2 
                              - \frac{1}{4} U_{yyx} \eta^2 - \frac{1}{4} U_{zzx} \zeta^2 \nonumber \\
                           & -\frac{1}{2} U_{xyx} \xi\eta - \frac{1}{2} U_{yzx} \eta\zeta
                               -\frac{1}{2} U_{zxx} \zeta\xi -\frac{1}{2} k_X \xi,  \nonumber \\
m_Y \ddot{y} = & -U_y -\frac{1}{4} U_{xxy} \xi^2 
                              - \frac{1}{4} U_{yyy} \eta^2 - \frac{1}{4} U_{zzy} \zeta^2 \nonumber \\
                           & -\frac{1}{2} U_{xyy} \xi\eta - \frac{1}{2} U_{yzy} \eta\zeta
                               -\frac{1}{2} U_{zxy} \zeta\xi -\frac{1}{2} k_Y \eta, \nonumber \\
m_Z \ddot{z} = & -U_z -\frac{1}{4} U_{xxz} \xi^2 
                              - \frac{1}{4} U_{yyz} \eta^2 - \frac{1}{4} U_{zzz} \zeta^2 \nonumber \\
                           & -\frac{1}{2} U_{xyz} \xi\eta - \frac{1}{2} U_{yzz} \eta\zeta
                               -\frac{1}{2} U_{zxz} \zeta\xi -\frac{1}{2} k_Z \zeta . 
\end{align}
This set of equations of motion 
may be derived from the potential
\begin{equation}
\label{ueff1}
U_{\text{eff}} (x,y,z) = U(x,y,z) + \frac{1}{4} k_X x\xi + 
\frac{1}{4} k_Y y\eta + \frac{1}{4} k_Z z\zeta , 
\end{equation}
which may be verified by 
direct differentiation.

\subsection{Application of the generalized pseudopotential
to the Paul-trap dynamics}
\label{secIIb}

Having presented the general analytical framework,
in this subsection we now apply our results to the special case
of three particles in the Paul trap.
In particular, we are interested in investigating
crystalline structures formed in the Paul trap.

\begin{figure}
\centering
\includegraphics[scale=0.6,angle=0]{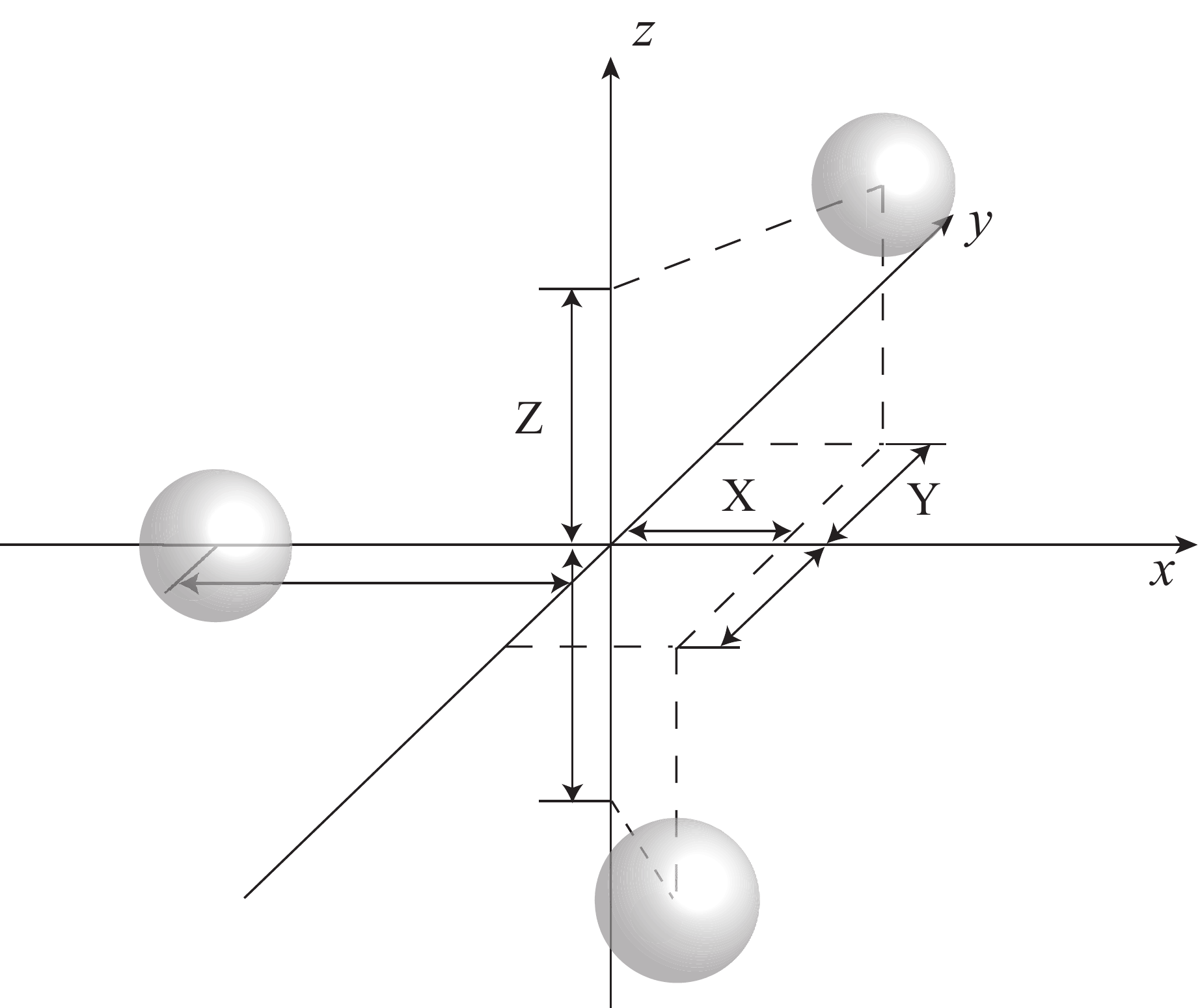}
\caption {\label{fig2}
Coordinate choice for the three-particle crystal formed in the Paul trap. 
Since the center of mass of the crystal is located at the center of the trap, 
knowing  $X,Y,Z,$ the coordinates of the top-most particle is enough to know the coordinates
of the other two particles.
}
\end{figure}

Any $N$-particle ion crystal, in particular the 
three-ion crystal we focus on in this section, 
is characterized by the geometric order in which the 
particles assemble themselves 
\cite{Nature,WSL,Kappler,Walth}, and, in  
addition, by the fact that this order is 
maintained during a micromotion cycle, during 
which the ions of a crystal execute a breathing motion 
around their equilibrium positions. 
Thus, due to this strong geometric correlation, 
knowing the coordinates $(X,Y,Z)$ of a {\it single} 
ion of the crystal, supplemented with the 
geometric configuration of the crystal, allows us to know the 
coordinates of all the other ions in the crystal. 
Fundamentally, 
therefore, the $N$-ion crystal problem is 
reduced to a single-particle problem. 
In the case of the three-ion crystal, 
the focus of this section, we choose 
the coordinates of the topmost ion 
(see box in Fig.~\ref{fig1}) to 
characterize the entire crystal. 
As shown in Fig.~\ref{fig2}, we denote its 
coordinates by $(X,Y,Z)$. Knowing 
$(X,Y,Z)$, together with the 
condition that the center of mass of this crystal 
is located at the center of the trap, is enough 
to know all 
coordinates $(X_i,Y_i,Z_i)$, $i=1,2,3$, of 
the three-ion crystal. 
 
%

Having chosen our coordinates,
we now derive equations of motion for three particles in a Paul trap.
We start with the Lagrangian of the system
\begin{equation}
\label{Lagrange}
\mathcal{L} = \left[T^{(1)}+T^{(2)}+T^{(3)}\right]
                        -\left[U^{(1)}_{\text{Trap}}+U^{(2)}_{\text{Trap}}+U^{(3)}_{\text{Trap}}\right]
                        -\left[U^{(1)-(2)}_{\text{Coul}}+U^{(2)-(3)}_{\text{Coul}}
                                +U^{(3)-(1)}_{\text{Coul}}\right],
\end{equation}
where, for $(X_i,Y_i,Z_i)$ and $(\dot{X}_i,\dot{Y}_i,\dot{Z}_i)$, 
the positions and velocities of the $i$th particle, respectively,
in suitable dimensionless units \cite{dim},  
\begin{equation}
\label{kin-i}
T^{(i)} = \frac{1}{2}(\dot{X}_i^2+\dot{Y}_i^2+\dot{Z}_i^2)
\end{equation}
denotes the kinetic energy of the $i$th particle,
\begin{equation}
\label{T-pot-i}
U^{(i)}_{\text{Trap}} = U_{\text{Trap}}(X_i,Y_i,Z_i) = 
\frac{1}{2} [ a+2q\cos(2\tau)] (X_i^2+Y_i^2-2Z_i^2) 
\end{equation}
denotes the trap potential 
of the $i$th particle, and
\begin{equation}
\label{C-pot-i}
U^{(i)-(j)}_{\text{Coul}} = \frac{1}{\sqrt{(X_i-X_j)^2+(Y_i-Y_j)^2+(Z_i-Z_j)^2}}
\end{equation}
denotes the Coulomb potential between
the $i$th and the $j$th particle.
Using the Euler-Lagrange equations \cite{LL,Euler},  
with (\ref{kin-i}), (\ref{T-pot-i}), 
and (\ref{C-pot-i}) in (\ref{Lagrange}), 
and relating $(X_i,Y_i,Z_i)$ to $(X,Y,Z)$ via 
the center-of-mass condition, 
we obtain a set of three equations of motion 
in each of the $x$, $y$, and $z$ directions, namely,
\begin{align}
6\ddot{X} &= -6X[a+2q\cos(2\tau)] + \frac{18X}{(9X^2+Y^2+Z^2)^{3/2}}, \nonumber \\
2\ddot{Y} &= -2Y[a+2q\cos(2\tau)] + \frac{Y}{2(Y^2+Z^2)^{3/2}} 
                                                            + \frac{2Y}{(9X^2+Y^2+Z^2)^{3/2}}, \nonumber \\
2\ddot{Z} &=+4Z[a+2q\cos(2\tau)] + \frac{Z}{2(Y^2+Z^2)^{3/2}}
                                                            + \frac{2Z}{(9X^2+Y^2+Z^2)^{3/2}},
\label{3ISYS} 
\end{align}
which, upon a comparison with (\ref{dynamic}), results in
$m_X = 6$, $m_Y = 2$, $m_Z = 2$, $\omega = 2$,
$k_X = 12q$, $k_Y = 4q$, $k_Z = -8q$, and
\begin{equation}
\label{u1}
U(X,Y,Z) = \frac{2}{(9X^2+Y^2+Z^2)^{1/2}} 
               + \frac{1}{2(Y^2+Z^2)^{1/2}} 
               + a\left[3X^2+Y^2-2Z^2\right].
\end{equation}
We note that $(X,Y,Z)$ in (\ref{u1}) may be replaced with $(x,y,z)$
upon taking a cycle average.
 
Eliminating the time-dependence from (\ref{3ISYS}) 
according to \cite{LL} to lowest order results in 
the pseudopotential 
\begin{equation}
\label{std-pseudo}
U_{\text{eff}}^{(\text{s})} = \frac{2}{\rho} + \frac{1}{2r} + \frac{1}{2}\mu_x^2x^2
					+ \frac{1}{2}\mu_y^2y^2 + \frac{1}{2}\mu_z^2z^2,
\end{equation}
where $\mu_x = [6(a+q^2/2)]^{1/2}$, $\mu_y = [2(a+q^2/2)]^{1/2}$, 
and $\mu_z = [4(q^2-a)]^{1/2}$. The pseudopotential (\ref{std-pseudo}) 
is not capable of reproducing all four crystal morphologies shown 
in the box of Fig.~\ref{fig1} and needs to be improved. 
In order to obtain the improved pseudopotential 
(\ref{ueff1}), capable of 
predicting all three-ion crystal morphologies, 
we need the partial derivatives of $U$ in (\ref{u1}). 
In particular,
\begin{align}
\label{partial-table}
U_{xx} &= 6a-\frac{18}{\rho^3}+\frac{486x^2}{\rho^5}, \nonumber \\
U_{yy} &= 2a-\frac{2}{\rho^3}+\frac{6y^2}{\rho^5} -\frac{1}{2r^3}+\frac{3y^2}{2r^5}, \nonumber \\
U_{zz} &= -4a-\frac{2}{\rho^3}+\frac{6z^2}{\rho^5}-\frac{1}{2r^3}+\frac{3z^2}{2r^5}, \nonumber \\
U_{xy} &= \frac{54xy}{\rho^5}, \nonumber \\
U_{yz} &= \frac{6yz}{\rho^5}+\frac{3yz}{2r^5}, \nonumber \\
U_{zx} &= \frac{54xz}{\rho^5},
\end{align}
where $\rho = (9x^2+y^2+z^2)^{1/2}$
and $r = (y^2+z^2)^{1/2}$,
which can be used to result in
\begin{equation}
\label{ueff2}
U_{\text{eff}} = \frac{2}{\rho}+\frac{1}{2r}+3ax^2+ay^2-2az^2
                          +\frac{3q x\xi\Delta + q y\eta\Delta - 2q z\zeta\Delta}{\Delta},
\end{equation}
where the last term on the right-hand side, while complicated,
may be evaluated explicitly
using (\ref{micro}) and (\ref{microD}) with the partial derivatives 
(\ref{partial-table}).
 
Our effective, generalized potential (\ref{ueff2}) allows for
an improved approximation to the pseudopotential 
beyond the lowest-order approximation (\ref{std-pseudo}) 
via expanding (\ref{ueff2}) up to 
first order in $1/\text{length}$.
Appendix~\ref{appB} shows the detailed steps
to arrive at the following improved 
three-ion pseudopotential:
\begin{align}
\label{ueff3}
\tilde{U}_{\text{eff}} &= \left[\frac{2}{\rho} + \frac{1}{2r}\right] + \left[\frac{3(2q^2+4a-a^2)}{4-a}\right]x^2
                                    +\left[\frac{2q^2+4a-a^2}{4-a}\right]y^2 + \left[\frac{2(2q^2-2a-a^2)}{2+a}\right]z^2
\nonumber \\
     &+\frac{q^2}{\rho^5}  \bigg[  \frac{324x^4}{(4-a)^2}+\frac{4y^4}{(4-a)^2}+\frac{4z^4}{(2+a)^2}
\nonumber \\
                                          &\qquad+9x^2 y^2 \left(
                                                                   \frac{8}{(4-a)^2}\right)  
\nonumber \\
                                          &\qquad-2y^2 z^2 \left(\frac{6}{(4-a)(2+a)}+\frac{1}{(4-a)^2}
                                                                  +\frac{1}{(2+a)^2}\right)
\nonumber \\
                                          &\qquad-18z^2 x^2 \left(\frac{6}{(4-a)(2+a)}+\frac{1}{(4-a)^2}
                                                                  +\frac{1}{(2+a)^2}\right) \bigg]
\nonumber \\
     &+\frac{q^2}{r^5} \left\{ \frac{y^4}{(4-a)^2}+\frac{z^4}{(2+a)^2} 
                                              -y^2z^2\left[\frac{1}{2(4-a)^2}+\frac{1}{2(2+a)^2}
                                                                  +\frac{3}{(4-a)(2+a)}\right] \right\}.
\end{align}
We will use this potential extensively in Sec.~\ref{secIV}. 
%
\section{Two-ion crystal morphology}
\label{secIII}

Before applying 
the formalism developed in Sec.~\ref{secII}
to the three-ion crystals formed in the Paul trap 
(see Sec.~\ref{secIV}), 
we review, in this section, in the spirit 
of a template, the two-ion case studied
in detail in Refs.~\cite{2ION,MB}. 
This allows us to demonstrate 
the power of the improved pseudopotential over
the standard pseudopotential in the most straightforward manner
in that the two-ion case is the simplest case that exhibits
crystal morphologies not predicted by the standard pseudopotential.

We start by pointing out the fact that a crystal in a Paul trap is formed
when the system is cooled, as particles find their respective, mutual,
simultaneous potential minima.
In terms of our improved pseudopotential 
this means that we may analytically determine 
the crystal configurations of the system by 
locating the minima of its pseudopotential. 
The potential minima may be obtained
straightforwardly by setting
(i) the first spatial derivatives of the potential equal to zero and
(ii) requiring that the second spatial derivatives of the potential are positive.
This methodology provides us with a technique 
that can be employed in general to investigate crystal morphologies
of a system for which the governing potential is well-known,
or approximately known, but sufficiently accurate. 
Furthermore, an additional bonus is the possibility to predict boundaries 
between different crystal structures, which are obtained
by tracking the stability of these minima in 
$(q,a)$ control-parameter space. 

\begin{figure}
\centering
\includegraphics[scale=0.27,angle=0]{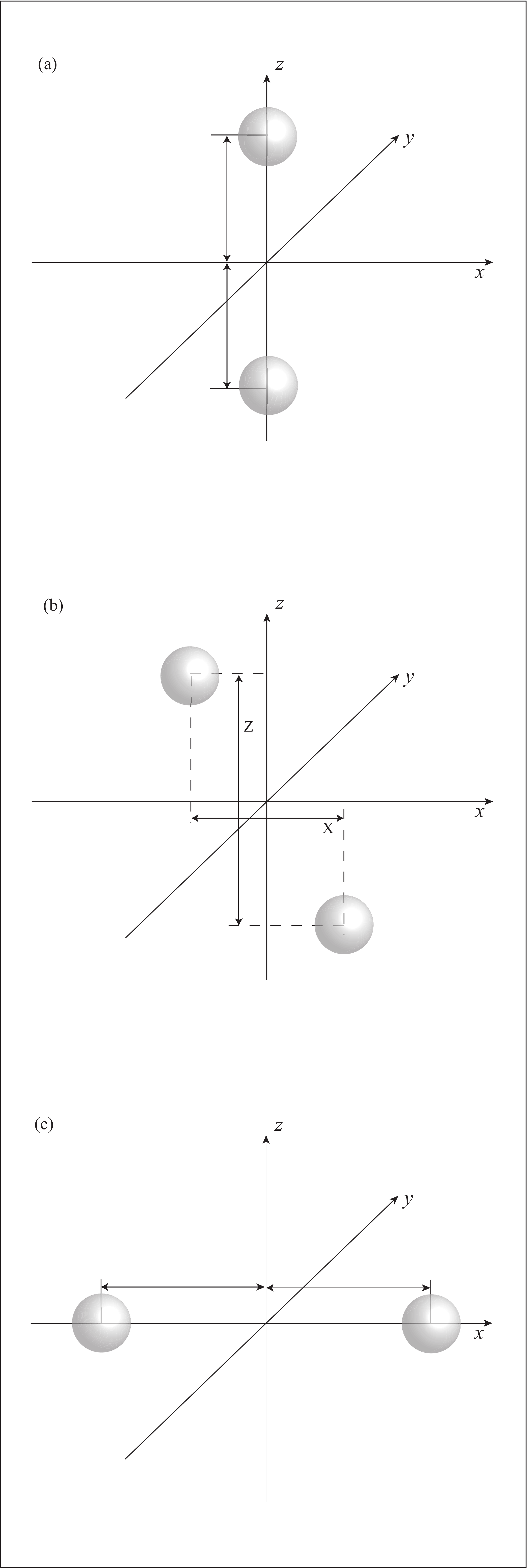}
\caption {\label{fig3}
Possible crystal configurations of two charged particles
stored in the Paul trap.
(a) shows a rod configuration, i.e., the particles are
lined up along the $z$-axis,
(b) shows a tilt configuration, i.e., the particles have
non-zero polar angle with respect to the $z$-axis, and
(c) shows a planar configuration, i.e.,
the particles form a horizontal rod in the $x$-$y$ plane
(here shown to lie along the $x$-axis).
Motivated by the fact that the tilt orientation in (b) 
is the most general among all three configurations,
we define the coordinates $X$ and $Z$ of the two-particle crystal, as shown, 
as the relative separation between the two ions 
along the $x$ and $z$ axes, respectively. Because of the cylindrical 
symmetry of the Paul trap we may, without restriction of 
generality, choose $Y\equiv 0$. 
}
\end{figure}

Applying the general pseudopotential derivation presented in 
Appendix~\ref{appA}
to the two-particle case, 
and closely following the steps presented in Sec.~\ref{secIIb}
(applied to the two-particle case), 
we obtain the effective potential for the two-particle system according to
\begin{equation}
\label{2Ueff}
U_{\text{eff}} = \frac{1}{r_2}
+\frac{1}{2}ax^2-az^2+\frac{qx\xi\Delta-2qz\zeta\Delta}{\Delta},
\end{equation}
where
\begin{align}
\xi =  [(m\omega^2-U_{zz})k_X x+U_{xz} k_Z z]/\Delta, \nonumber \\
\zeta = [(m\omega^2-U_{xx})k_Z z+U_{xz} k_X x]/\Delta, \nonumber \\
\Delta = (m\omega^2-U_{xx})(m\omega^2-U_{zz})-U_{xz}^2,
\end{align}
and
\begin{equation}
U_{xx} = a-\frac{1}{r_2^3}+\frac{3x^2}{r_2^5}, \qquad
U_{xz} = \frac{3xz}{r_2^5}, \qquad
U_{zz} = -2a-\frac{1}{r_2^3}+\frac{3z^2}{r_2^5},
\end{equation}
where $r_2=(x^2+z^2)^{1/2}$.
Upon comparison with the two-coordinate formulation
(see Appendix~\ref{appA}),
this results in $k_X = 2q$, $k_Z = -4q$, $m= 1$, $\omega = 2$.
Our choice of coordinates $x$ and $z$ (with time averaging),
along with the observed crystal orientations, 
is presented in Fig.~\ref{fig3}. The potential 
(\ref{2Ueff}) may be compared with 
the standard pseudopotential
for the two-particle case \cite{2ION,MB}
\begin{equation}
\label{2Ueffs}
U_{\text{eff}}^{(s)} = \frac{1}{r_2}+\frac{1}{2}\left(a+\frac{q^2}{2}\right)x^2
+(q^2-a)z^2.
\end{equation}
Equipped with the two potentials (\ref{2Ueff}) and (\ref{2Ueffs}),
we are now ready to determine the potential minima by
following steps (i) and (ii) as defined above
to obtain the different morphology boundaries predicted by
the two different pseudopotentials, thus demonstrating
the power of the improved, generalized pseudopotential.
\begin{itemize} 
\item{Rod-tilt boundary.}
In order to extract the rod-tilt boundary, we first
assume a rod-configuration, i.e., $x = 0$, and
investigate the critical points of the effective potentials
(\ref{2Ueff}) and (\ref{2Ueffs}) in $z$. Then, we investigate the
stability of the two potentials in $x$. 
A transition from a stable to an unstable potential landscape
in the $x$-direction corresponds to 
a rod$\leftrightarrow$tilt crystal metamorphosis.
The boundary of the 
metamorphosis satisfies the two conditions 
\begin{equation}
\label{2rtf}
\frac{1}{z}\frac{\partial U_{\text{eff}}(x=0,z)}{\partial z} =0
\end{equation}
and
\begin{equation}
\label{2rts}
\frac{\partial^2 U_{\text{eff}}(x,z)}{\partial x^2}\bigg|_{x=0} = 0. 
\end{equation}
Equivalent expressions hold 
for the standard pseudopotential $U_{\text{eff}}^{(s)}$.
Numerically 
solving the simultaneous equations 
(\ref{2rtf}) and (\ref{2rts}) \cite{NumRec}, 
we obtain the two-particle rod-tilt boundary predicted by
the generalized pseudopotential.
The standard pseudopotential counterparts of 
(\ref{2rtf}) and (\ref{2rts}) can be solved 
exactly and analytically, which results in the following 
simple analytical expression for the boundary line, i.e.,
\begin{equation}
\label{2bound-srt}
a = f^{(N=2)}_{\text{rod}\leftrightarrow \text{tilt}} = \frac{1}{2}q^2.
\end{equation}
As will be shown below, the boundary (\ref{2bound-srt}) predicts 
a degenerate case of two-particle morphologies,
reflecting the inability of the standard pseudopotential
to predict tilted two-ion crystals.
%
\item{Tilt-planar boundary.}
This time, we assume a planar configuration,
i.e., $z = 0$, and investigate the critical points
of the effective potentials (\ref{2Ueff}) and (\ref{2Ueffs}) in $x$.
Then, we investigate the stability of the potentials
in $z$, where we look for the transition from
stable to unstable configurations.
The generalized pseudopotential is then
subjected to the following two conditions:
\begin{equation}
\label{2tpf}
\frac{1}{x}\frac{\partial U_{\text{eff}}(x,z=0)}{\partial x} =  0
\end{equation}
and
\begin{equation}
\label{2tps}
\frac{\partial^2 U_{\text{eff}}(x,z)}{\partial z^2}\bigg|_{z=0} = 0.
\end{equation}
Solving (\ref{2tpf}) and (\ref{2tps}) results in
the generalized pseudopotential prediction of the boundary line
between the tilt and planar phases of two-particle crystals.
Repeating the same procedure for the standard pseudopotential,
which, once again, may be solved analytically,  
results in the boundary
\begin{equation}
\label{2bound-stp}
a = f^{(N=2)}_{\text{tilt} \leftrightarrow \text{planar}} = \frac{1}{2} q^2.
\end{equation}
\end{itemize} 
 
\begin{figure}
\includegraphics[scale=2.2,angle=0]{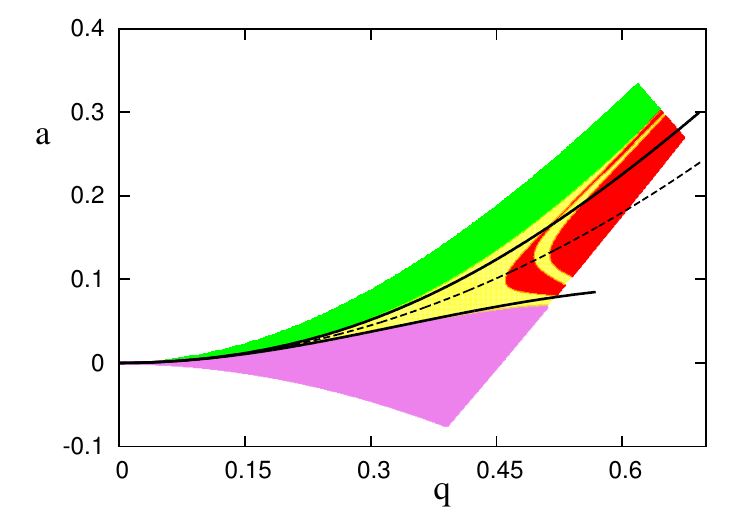}
\caption {\label{fig4}
(Color Online)  Two-ion crystal-configuration map
in $(q,a)$ control-parameter space, obtained 
from numerical solutions of (\ref{TDEM3}). 
The green region is the rod phase,
yellow regions are the tilt phase, and the
purple region is the planar phase.
Red regions denote a non-crystal region,
where the two ions are stably trapped in the Paul trap
but never form a crystal \cite{EMB}.  
The dashed line denotes the transition boundaries
(\ref{2bound-srt}) and (\ref{2bound-stp})
predicted by the standard pseudopotential (\ref{2Ueffs}).
The solid lines denote the transition boundaries
predicted by the generalized pseudopotential (\ref{2Ueff}). 
Top curve: Rod $\leftrightarrow$ tilt boundary; 
bottom curve: Tilt $\leftrightarrow$ planar boundary.
}
\end{figure}

The two boundaries (\ref{2bound-srt}) and (\ref{2bound-stp})
are identical. Thus, according to the standard pseudopotential, 
the tilt morphology is degenerate, existing only on the 
line $a=q^2/2$. As we will see shortly (see Fig.~\ref{fig4}) this 
prediction is incorrect. 
 
Figure~\ref{fig4} shows our numerical simulation results
of the crystal orientations of the two-ion crystals in the Paul trap.
Also shown are the boundary lines derived above
[see (\ref{2bound-srt}) and (\ref{2bound-stp}) 
for the standard pseudopotential case].
We see that, as pointed out in \cite{2ION,MB},
the standard pseudopotential fails to predict the existence
of the tilt-morphology, i.e., according to the 
standard pseudopotential
this crystal phase exists only on the zero-width curve $a=q^2/2$.
In fact, according to the standard pseudopotential (\ref{2Ueffs}),
the case $a=q^2/2$ results in a radially symmetric potential,
corresponding to a degenerate crystal-configuration state, where
all radially symmetric morphologies (rod, tilt, and planar) are possible,
but are only neutrally stable (the second derivative of the pseudopotential is zero).
Thus the boundary lines predicted on the basis of the standard pseudopotential 
do not agree well with the exact, numerical results.
In contrast, the generalized, improved pseudopotential correctly 
predicts the existence of 
all three morphologies, including the fact 
that they all have finite areas in $(q,a)$ control-parameter 
space. 
In addition, for $q\lesssim 0.45$, the predicted 
morphology boundaries are in excellent agreement with
our simulations.
This is remarkable since the additional terms we keep in 
the generalized pseudopotential are not only improving the 
accuracy, but contain additional physics not contained in 
the lower-order standard pseudopotential. Thus, 
the generalized pseudopotential captures necessary, important aspects
of crystal morphology observed in the Paul-trap system.

\begin{figure}
\includegraphics[scale=1.4,angle=0]{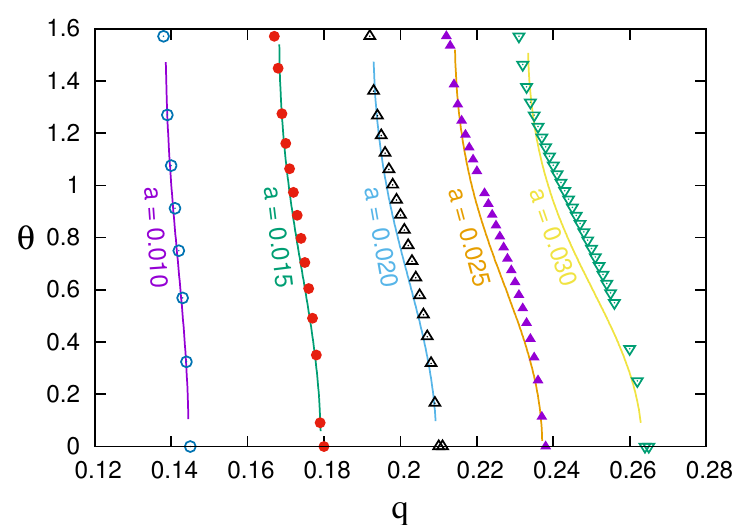}
\caption {\label{fig5}
(Color Online)  
Polar angle $\theta$ of the two-ion tilt phase observed in the Paul trap
as a function of $q$, obtained from 
numerical simulations of (\ref{TDEM3}). 
Shown are the cases 
$a=0.01$ (open circles),
0.015 (filled circles), 
0.02 (open triangles), 
0.025 (filled triangles), 
and 0.03 (open, inverted triangles).
Solid lines are the $\theta$ solutions of the transcendental equation 
(\ref{2tilt-3}). 
}
\end{figure}

A new result, further illustrating the power of 
the improved pseudopotential, is the prediction
of the polar angle of the tilt phase, i.e., 
the angle that the tilted
two-ion rod crystal forms with respect to the $z$-axis.
To this end, we approximate the generalized,
improved pseudopotential in (\ref{2Ueff}) up to $1/$length,
following the same methodology as described in detail in Appendix~\ref{appB}.
The result is
\begin{align}
\tilde{U}_{\text{eff}} &= \frac{1}{r_2} + \frac{2q^2+4a-a^2}{2(4-a)}x^2
+\frac{2q^2-2a-a^2}{2+a}z^2 \nonumber \\
&+\frac{q^2}{r_2^5}\left[ \frac{2x^4}{(4-a)^2}
+\frac{2z^4}{(2+a)^2}-\frac{4(17+2a-a^2)x^2z^2}{(4-a)^2(2+a)^2} \right].
\end{align}
Assuming now that we are in the tilt phase $(x\neq 0; z\neq 0)$,
the polar angle is obtained as the solution of 
the following two simultaneous equations: 
\begin{align}
\label{2tilt-1}
\frac{1}{x}\frac{\partial\tilde{U}_{\text{eff}}}{\partial x} =& 
-\frac{1}{r_2^3} + \frac{2q^2+4a-a^2}{4-a} 
\nonumber \\
&-\frac{5q^2}{r_2^7}
\left[\frac{2x^4}{(4-a)^2}+\frac{2z^4}{(2+a)^2}-\frac{4(17+2a-a^2)x^2z^2}{(4-a)^2(2+a)^2}\right] 
\nonumber \\
&+\frac{q^2}{r_2^5}\left[ \frac{8x^2}{(4-a)^2}-\frac{8(17+2a-a^2)z^2}{(4-a)^2(2+a)^2}\right] 
=0
\end{align}
and
\begin{align}
\label{2tilt-2}
\frac{1}{z}\frac{\partial\tilde{U}_{\text{eff}}}{\partial z} = &
-\frac{1}{r_2^3} +\frac{2(2q^2-2a-a^2)}{2+a}
\nonumber \\
&-\frac{5q^2}{r_2^7}
\left[\frac{2x^4}{(4-a)^2}+\frac{2z^4}{(2+a)^2}-\frac{4(17+2a-a^2)x^2z^2}{(4-a)^2(2+a)^2}\right] 
\nonumber \\
&+\frac{q^2}{r_2^5}\left[ \frac{8z^2}{(2+a)^2}-\frac{8(17+2a-a^2)x^2}{(4-a)^2(2+a)^2}\right] 
=0.
\end{align}
To obtain the solution, we 
subtract (\ref{2tilt-2}) from (\ref{2tilt-1}) to obtain 
\begin{equation}
\label{2tilt-3}
\omega^2_x-\omega^2_z
+\frac{q^2}{r_2^3}\left\{
\frac{8\sin^2(\theta)}{(4-a)^2}-\frac{8\cos^2(\theta)}{(2+a)^2}
-\frac{8(17+2a-a^2)}{(4-a)^2(2+a)^2}[\cos^2(\theta)-\sin^2(\theta)]\right\}=0,
\end{equation}
where $\theta$ is the polar angle, 
\begin{align}
\omega^2_x &= \frac{2q^2+4a-a^2}{4-a},  \nonumber \\
\omega^2_z &= \frac{2(2q^2-2a-a^2)}{2+a},
\end{align}
$x = r_2\sin(\theta)$, and $z = r_2\cos(\theta)$. 
Adding (\ref{2tilt-1}) and (\ref{2tilt-2}) yields
\begin{align}
\label{2tilt-4}
\frac{1}{r_2^3}
= &(\omega^2_x+\omega^2_z)\left\{
2 - q^2\left[ \frac{8\sin^2(\theta)}{(4-a)^2}+\frac{8\cos^2(\theta)}{(2+a)^2}
-\frac{8(17+2a-a^2)}{(4-a)^2(2+a)^2}\right]\right. \\ \nonumber
& \left.+ 10q^2\left[ \frac{2\sin^4(\theta)}{(4-a)^2}
+\frac{2\cos^4(\theta)}{(2+a)^2}-\frac{4(17+2a-a^2)\cos^2(\theta)\sin^2(\theta)}
{(4-a)^2(2+a)^2}\right]\right\}^{-1},
\end{align}
which may now be inserted into (\ref{2tilt-3}) to result in
a transcendental equation for $\theta$.
Figure~\ref{fig5} shows the numerical solutions 
of the resulting transcendental equation together with 
the polar angles $\theta$ obtained from our numerical simulations 
of (\ref{TDEM3}).
The agreement between our analytical predictions and the 
numerical simulations is excellent.
This shows the remarkable accuracy of the generalized pseudopotential method, 
which is possible only because the 
generalized pseudopotential contains hidden physics
beyond the standard approach.

%
\section{Three-ion crystal morphology}
\label{secIV}

Encouraged by the success 
of the improved pseudopoential
for the two-ion case illustrated in Sec.~\ref{secIII},
in this section we now investigate the case of 
three particles simultaneously stored 
in the Paul trap, using the improved pseudopotential.
The improved, generalized pseudopotential $U_{\text{eff}}$ and
its $1/\text{length}$ approximated version $\tilde{U}_{\text{eff}}$ 
[see (\ref{ueff2}) and (\ref{ueff3}), respectively]
for the three particles in the Paul trap are much more complicated
than the standard potential $U_{\text{eff}}^{(\text{s})}$ in (\ref{std-pseudo}).
A natural question to ask is what additional
physics and insight we gain from these more 
complicated expressions.
The answer is that, in analogy to the two-ion case, 
the improved pseudopotential predicts
a new regime of crystal morphology that is not captured
in the standard approach. 

\subsection{Standard vs. generalized pseudopotential prediction}
\label{secIVa}

\begin{figure}
\includegraphics[scale=2.2,angle=0]{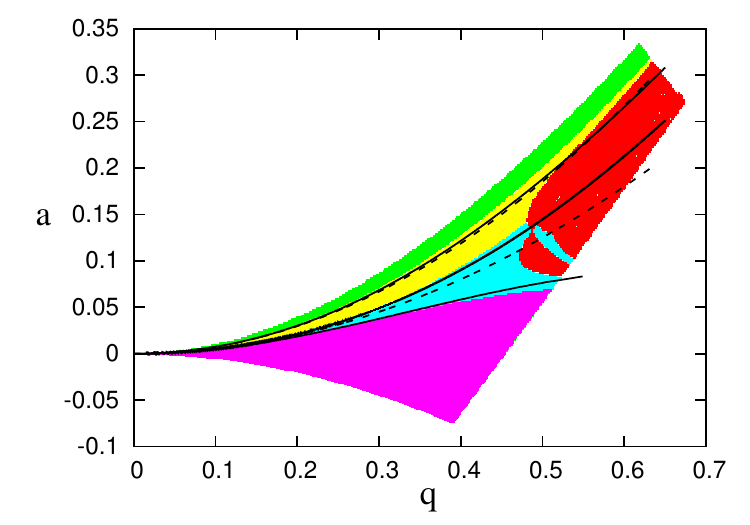}
\caption {\label{fig6}
(Color Online)  Standard and improved pseudopotential predictions
for three-particle morphology transitions in $(q,a)$ 
control-parameter space. 
Green, yellow, blue, and magenta regions denote
rod, pop-out, tilt, and planar crystal morphologies, respectively.
The red region indicates a region in $(q,a)$ 
control-parameter space where no stable three-ion crystals exist.
The top dashed line is the standard
pseudopotential prediction 
for the rod $\leftrightarrow$ pop-out transition. 
The standard pseudopotential (incorrectly) predicts 
that the pop-out $\leftrightarrow$ tilt,
and tilt $\leftrightarrow$ planar transitions are 
degenerate (represented by the bottom dashed line). 
Solid lines (top to bottom) 
correspond to the correct predictions 
of the generalized pseudopotential for 
rod $\leftrightarrow$ pop-out, pop-out $\leftrightarrow$ tilt,
and tilt $\leftrightarrow$ planar transitions. 
}
\end{figure}

To start, we first investigate the standard potential 
$U^{(\text{s})}_{\text{eff}}$ in (\ref{std-pseudo}).
We recall that, to investigate the boundary lines
between different crystal morphologies, we need
both the first and second derivatives of the potential 
to be equal to zero. The results for the first derivatives are: 
\begin{align}
\label{fderiv}
\frac{\partial U_{\text{eff}}^{(\text{s})}}{\partial x} 
&= \left(-\frac{18}{\rho^3}+\mu_x^2\right)x = 0, \nonumber \\
\frac{\partial U_{\text{eff}}^{(\text{s})}}{\partial y}
&= \left(-\frac{2}{\rho^3}-\frac{1}{2r^3}+\mu_y^2\right)y = 0, \nonumber \\
\frac{\partial U_{\text{eff}}^{(\text{s})}}{\partial z}
&= \left(-\frac{2}{\rho^3}-\frac{1}{2r^3}+\mu_z^2\right)z = 0.
\end{align}
The results for the 
second derivatives, including mixed derivatives, are: 
\begin{align}
\label{sderiv}
\frac{\partial^2U_{\text{eff}}^{(\text{s})}}{\partial x^2}
&=18\frac{18x^2-y^2-z^2}{\rho^5}+\mu_x^2, \nonumber \\
\frac{\partial^2U_{\text{eff}}^{(\text{s})}}{\partial y^2}
&=2\frac{-9x^2+2y^2-z^2}{\rho^5}+\frac{2y^2-z^2}{2r^5}+\mu_y^2, \nonumber \\
\frac{\partial^2U_{\text{eff}}^{(\text{s})}}{\partial z^2}
&=2\frac{-9x^2-y^2+2z^2}{\rho^5}+\frac{2z^2-y^2}{2r^5}+\mu_z^2, \nonumber \\
\frac{\partial^2U_{\text{eff}}^{(\text{s})}}{\partial x \partial y}
&=\frac{54xy}{\rho^5}, \nonumber \\
\frac{\partial^2U_{\text{eff}}^{(\text{s})}}{\partial y \partial z}
&=\frac{6yz}{\rho^5}+\frac{3yz}{2r^5}, \nonumber \\
\frac{\partial^2U_{\text{eff}}^{(\text{s})}}{\partial z \partial x}
&=\frac{54zx}{\rho^5}.
\end{align}
Because of its structure the solutions 
of the nonlinear system of 
equations (\ref{fderiv}) can be classified into 
the following eight distinct cases:
\begin{itemize}
\item case 1: $(x = 0,y = 0,z = 0)$  - Coulomb repulsion, not viable;
\item case 2: $(x = 0,y = 0,z\neq0)$ - Rod phase, observed;
\item case 3: $(x = 0,y\neq0,z = 0)$ - Preferred direction in $x$-$y$ plane, not viable;
\item case 4: $(x = 0,y \neq 0, z \neq 0)$ - Tilted rod phase, not observed;
\item case 5: $(x\neq0,y=0,z=0)$ - Preferred direction in $x$-$y$ plane, not viable;
\item case 6: $(x\neq0,y=0,z\neq0)$ - Pop-out phase, observed;
\item case 7: $(x\neq0,y\neq0,z=0)$ - Planar phase
                        [symmetry$\rightarrow$ $y = \sqrt{3}x$], observed;
\item case 8: $(x\neq0,y\neq0,z\neq0)$ - Tilt phase, observed.
\end{itemize}
We now investigate cases 2, 6, and 7 in detail, which yields the 
morphology boundaries between all four observed cases, i.e., 
cases 2, 6, 7, and 8. 
\begin{enumerate}
\item Rod phase (case 2)
 
This case yields $z^3 = 5/[8(q^2-a)]$.
Inserting it into (\ref{sderiv}) results in 
zero for all mixed derivatives, and
positive for the second derivative in $z$.
Stability is lost in the $x$ and $y$ directions, respectively, when
\begin{align}
\label{c2bound}
\frac{\partial^2U_{\text{eff}}^{(\text{s})}}{\partial x^2} = 0 \rightarrow a = \frac{43}{58} q^2, \nonumber \\
\frac{\partial^2U_{\text{eff}}^{(\text{s})}}{\partial y^2} = 0 \rightarrow a = \frac{1}{2} q^2,
\end{align}
which indicates that the tilted rod phase (case 4), the case where the stability is lost in $y$,
does not occur, because the pop-out phase (case 6), where the stability is lost in $x$,
occurs first, before case 4 can be realized. 

\item Pop-out phase (case 6)

This case yields $\rho(y=0) = \sqrt{9x^2+z^2} = [3/(a+q^2/2)]^{1/3}$ 
and $z = [3/(22q^2-28a)]^{1/3}$.
Inserting these results into (\ref{sderiv}) results in 
zero for the $x,y$ and $y,z$ mixed derivatives
(and positive for the $z,x$ mixed derivative
and second derivatives in $x$ and $z$). 
Therefore,  
stability for this case 
is lost in $y$ direction,
resulting in the tilt-phase (case 8), when
\begin{equation}
\label{c6bound}
\frac{\partial^2U_{\text{eff}}^{(\text{s})}}{\partial y^2} = 0 \rightarrow a = \frac{1}{2} q^2.
\end{equation}

\item Planar phase (case 7)

This case yields $y = \sqrt{3}x = [3/(a+q^2/2)]^{1/3}/2$.
Inserting this result into (\ref{sderiv}) results in 
zero for the $y,z$ and $z,x$ mixed derivatives
(and positive for the $x,y$ mixed derivative
and second derivatives in $x$ and $y$).
Therefore, stability for this case 
is lost in $z$ direction,
resulting in the tilt phase (case 8), when
\begin{equation}
\label{c7bound}
\frac{\partial^2U_{\text{eff}}^{(\text{s})}}{\partial z^2} = 0 \rightarrow a = \frac{1}{2} q^2.
\end{equation}
\end{enumerate}
Comparing the case-6 and the case-8 results, 
we conclude that the tilt phase,
under the conventional, standard pseudopotential, exists 
only on the line $a=q^2/2$ in
the $(q,a)$ control-parameter space.
The boundary lines corresponding to
the transitions between the four observed 
crystal morphologies (cases 2, 6, 7, and 8) 
may be found in Fig.~\ref{fig6}. 

We now move on to the predictions of the generalized pseudopotential.
Starting, this time, with (\ref{ueff2}), we obtain the boundaries between
rod and pop-out, pop-out and tilt, and tilt and planar phases by
following exactly the same methodology employed above 
for the standard pseudopotential,
i.e., for the rod and pop-out boundary via eliminating $z$ from 
\begin{equation}
\label{num-bound1}
\frac{\partial    U_{\text{eff}}(x=0,y=0,z)}{\partial z} = 0, \qquad
\frac{\partial^2U_{\text{eff}}(x,y=0,z)}{\partial x^2}\bigg|_{x=0} = 0,
\end{equation}
for the pop-out and tilt boundary via eliminating $x$ and $z$ from 
\begin{equation}
\label{num-bound2}
\frac{\partial    U_{\text{eff}}(x,y=0,z)}{\partial z} = 0, \qquad
\frac{\partial    U_{\text{eff}}(x,y=0,z)}{\partial x} = 0, \qquad
\frac{\partial^2U_{\text{eff}}(x,y,z)}{\partial y^2}\bigg|_{y=0} = 0,
\end{equation}
and for the tilt and planar boundary via eliminating $x$ from 
\begin{equation}
\label{num-bound3}
\frac{\partial    U_{\text{eff}}(x,y=\sqrt{3}x,z=0)}{\partial x} = 0, \qquad
\frac{\partial^2U_{\text{eff}}(x,y=\sqrt{3}x,z)}{\partial z^2}\bigg|_{z=0} = 0. 
\end{equation}
The analytical computation of explicit analytical solutions of the boundaries
is not straightforward. Therefore, we compute the 
solutions of (\ref{num-bound1}), 
(\ref{num-bound2}), and (\ref{num-bound3}) numerically,  
which results in the three solid lines  
in Fig.~\ref{fig6}, representing 
the three boundaries 
between the four three-particle crystal morphologies.
The agreement between the boundary lines 
between the four different crystal morphologies,  
determined as a result of our numerical simulations 
of (\ref{TDEM3}) (green, yellow, blue, and magenta regions 
in Fig.~\ref{fig6}), and 
the boundary lines predicted by the generalized improved pseudopotential
is excellent.
The generalized improved pseudopotential,
as in the two-particle case 
discussed in Sec.~\ref{secIII}, in contrast to the standard pseudopotential, 
also predicts both the existence and the location 
of the new three-particle tilt phase correctly.

\subsection{$1/$length approximated generalized pseudopotential}
\label{secIVb}

In this subsection, closely following the steps in 
Sec.~\ref{secIVa}, 
we investigate the three-ion crystal
in a Paul trap with the help of the $1/\text{length}$ approximated,
generalized pseudopotential $\tilde{U}_{\text{eff}}$, 
defined in (\ref{ueff3}). In the absence of simple analytical expressions, 
the motivation behind this approach is to obtain a 
better analytical handle
on the generalized, improved pseudopotential approach. 

As will be shown below, while still somewhat complicated,
the $1/$length approximation yields 
tractable analytical expressions for 
the pseudopotential minima and their stability, 
and thus we are able to
determine all boundaries fully analytically.

\begin{figure}
\includegraphics[scale=2.2,angle=0]{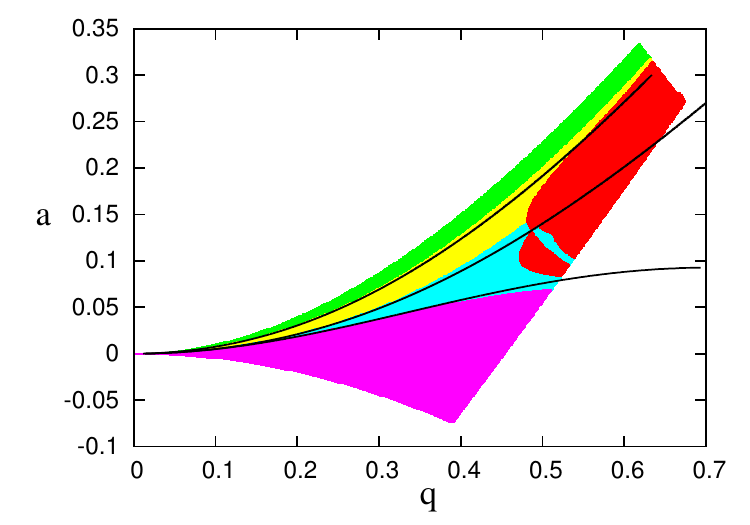}
\caption {\label{fig7}
(Color Online)  Improved, 1/length approximated 
generalized three-ion pseudopotential prediction
of crystal morphology transitions in ($q,a$) control-parameter space.
Colored areas code for the four observed three-particle crystal morphologies. 
They are imported from Fig.~\ref{fig6} without any modifications. 
Solid lines are the analytical predictions
of the boundary lines between the four 
three-particle crystal morphologies as computed 
in Sec.~\ref{secIVb}.
}
\end{figure}

\begin{enumerate}
\item Rod $\rightarrow$ pop-out boundary

Since we start in the rod phase, we have 
$x = y = 0$ in this case, and 
we need only
\begin{equation}
\label{c2z}
\frac{1}{z}\frac{\partial \tilde{U}_{\text{eff}}(x=0,y=0,z)}{\partial z} = 
\left[-\frac{2}{z^3}-\frac{1}{2z^3}\right] +
\frac{4(2q^2-2a-a^2)}{2+a} -
\frac{q^2}{z^3}\left[\frac{5}{(2+a)^2}\right].
\end{equation}
Equating (\ref{c2z}) to zero results in
\begin{equation}
\label{zc2}
z = \left[ \frac{10q^2+5(2+a)^2}{8(2+a)(2q^2-2a-a^2)} \right]^{\frac{1}{3}}.
\end{equation}

From the standard pseudopotential case we know
that the stability of the rod configuration is 
lost in $x$-direction.
In order to obtain the morphology boundary to the pop-out phase (case 6),
we now evaluate
\begin{align}
\frac{\partial^2\tilde{U}_{\text{eff}}(x,y=0,z)}{\partial x^2} \Bigg|_{x=0} = 
&-\frac{18}{z^3}+\frac{6(2q^2+4a-a^2)}{4-a} 
-\frac{45q^2}{z^3}\left[\frac{4}{(2+a)^2}\right] \nonumber \\
&-\frac{q^2}{z^3}\left[
\frac{144(17+2a-a^2)}{(4-a)^2(2+a)^2}\right],
\end{align}
which, upon equating it to 0, together with (\ref{zc2}), results in
\begin{align}
&q^4 (-116a^2+3112a-14048) + q^2 (-1328a^3+4728a^2+13392a-2752) \nonumber \\
+&(29a^6-58a^5-580a^4+232a^3+3712a^2+3712a)= 0.
\end{align}
This equation can be solved for $q^2$ using the quadratic 
solution formula. The result is shown as the 
top solid line 
in Fig.~\ref{fig7}.

\item Pop-out $\rightarrow$ tilt boundary

Starting in 
the pop-out phase, we have two degrees of freedom,
namely, $x$ and $z$.
Therefore, we now need
\begin{align}
\label{Vb2-1}
\frac{1}{x}\frac{\partial \tilde{U}_{\text{eff}}(x,y=0,z)}{\partial x} =  &
-\frac{6(2q^2+4a-a^2)}{4-a} - \frac{18}{(9x^2+z^2)^{3/2}} \nonumber \\
&+ \frac{144q^2[9(2+a)^2x^2-(17+2a-a^2)z^2]}{(4-a)^2(2+a)^2(9x^2+z^2)^{5/2}} \nonumber \\
&-\frac{180q^2[81(2+a)^2x^4-18(17+2a-a^2)x^2z^2+(4-a)^2z^4]}{(4-a)^2(2+a)^2(9x^2+z^2)^{7/2}}
\end{align}
and
\begin{align}
\label{Vb2-2}
\frac{1}{z}\frac{\partial \tilde{U}_{\text{eff}}(x,y=0,z)}{\partial z} = &
-\frac{1}{2z^3}-\frac{q^2}{(2+a)^2z^3}+\frac{4(2q^2-2a-a^2)}{2+a}
-\frac{2}{(9x^2+z^2)^{3/2}} \nonumber \\ &
+\frac{16q^2[-9(17+2a-a^2)x^2+(4-a)^2z^2]}{(4-a)^2(2+a)^2(9x^2+z^2)^{5/2}} \nonumber \\ &
-\frac{20q^2[81(2+a)^2x^4-18(17+2a-a^2)x^2z^2+(4-a)^2z^4]}{(4-a)^2(2+a)^2(9x^2+z^2)^{7/2}}
\end{align}
for the first derivatives, and
\begin{align}
\label{Vb2-3}
\frac{\partial^2\tilde{U}_{\text{eff}}(x,y,z)}{\partial y^2} \Bigg|_{y=0} = &
\frac{2(2q^2+4a-a^2)}{4-a} - \frac{1}{2z^3} - \frac{5q^2}{(2+a)^2z^3} 
- \frac{4(17+2a-a^2)q^2}{(4-a)^2(2+a)^2z^3} \nonumber \\ &
- \frac{2}{(9x^2+z^2)^{3/2}} 
+\frac{16q^2[9(2+a)^2x^2-(17+2a-a^2)z^2]}{(4-a)^2(2+a)^2(9x^2+z^2)^{5/2}} \nonumber \\ &
-\frac{20q^2[81(2+a)^2x^4-18(17+2a-a^2)x^2z^2+(4-a)^2z^4]}{(4-a)^2(2+a)^2(9x^2+z^2)^{7/2}}
\end{align}
for the second derivative. 
The transition from the pop-out phase
to the tilt-phase is observed when
the pop-out crystal loses its stability in $y$-direction.

Now, subtracting 9 times (\ref{Vb2-3}) from (\ref{Vb2-1}) results in
\begin{equation}
\label{Vb2-4}
z = \left\{\frac{3}{8(2q^2+4a-a^2)} \left[ (4-a) + \frac{2q^2}{4-a} + \frac{72q^2}{(2+a)^2} \right] \right\}^{1/3}.
\end{equation}
Subtracting (\ref{Vb2-1}) from 9 times (\ref{Vb2-2}) and
rearranging terms results in
\begin{equation}
\label{Vb2-5}
\delta(9x^2+z^2)^{5/2} = \beta x^2 + \gamma,
\end{equation}
where 
\begin{align} 
\label{Vb2-6}
& \delta = \frac{9}{2z^3} + \frac{9q^2}{(2+a)^2z^3} + 42a 
              + 12q^2\left[\frac{1}{4-a} - \frac{6}{2+a}\right], \nonumber \\
& \beta = -\frac{3888(7+2a)q^2}{(4-a)^2(2+a)^2}, \nonumber \\
& \gamma = \frac{432(11-2a)q^2z^2}{(4-a)^2(2+a)^2}.
\end{align}
At this point, together with $z$ in (\ref{Vb2-4}), we realize that 
(\ref{Vb2-5}) is a quintic equation in $x^2$. A
detailed discussion of how to 
obtain the desired
root may be found in Appendix~\ref{appC}. 
We find
\begin{equation}
\label{Vb2-7}
x = z\sqrt{\frac{11-2a}{9(7+2a)}} ,
\end{equation}
which, together with (\ref{Vb2-4}), 
may now be used to result in a single analytical expression
that relates $q$ and $a$, which yields the pop-out 
$\rightarrow$ tilt boundary line:
\begin{align}
\label{Vb2-8}
&3(4-a)^3a(2+a)^3 
-8(4-a)(2+a)(12-92a+17a^2)q^2 \nonumber \\ 
&-4(1064-580a+86a^2-3a^3)q^4= 0.
\end{align}
Using the quadratic solution formula, we may now solve
for $q^2$. The resulting boundary is shown in Fig.~\ref{fig7} 
as the second solid line from the top. 

\item Planar $\rightarrow$ tilt boundary

In order to access the third and last boundary, we evaluate
\begin{align}
\label{Vb3-1}
\frac{1}{x}\frac{\partial \tilde{U}_{\text{eff}}(x=0,y=\sqrt{3}x,z=0)}{\partial x} = &
-\frac{\sqrt{3}}{2x^3}-\frac{\sqrt{3}q^2}{(4-a)^2x^3}+\frac{12(2q^2+4a-a^2)}{4-a}
\end{align}
and
\begin{align}
\label{Vb3-2}
\frac{\partial^2\tilde{U}_{\text{eff}}(x,y=\sqrt{3}x,z)}{\partial z^2} \Bigg|_{z=0} = &
-\frac{\sqrt{3}}{12x^3}+\frac{4(2q^2-2a-a^2)}{2+a}-\frac{\sqrt{3}q^2}{2(4-a)^2x^3} \nonumber \\ &
+\frac{\sqrt{3}q^2}{3(2+a)^2x^3}-\frac{18\sqrt{3}q^2}{(4-a)^2(2+a)^2x^3} . 
\end{align}
Setting both (\ref{Vb3-1}) and (\ref{Vb3-2}) equal to zero, then
eliminating $x$, we obtain the $1/$length approximated
improved pseudopotential prediction of the tilt-planar boundary 
of three particles stored in the Paul trap according to 
\begin{align}
a(2+a)^2(4-a)^3 + 8(4-a)(4 -9a-4a^2)q^2-4(24+8a+a^2)q^4 = 0, 
\end{align}
which, as in case 2 above, may be solved for $q^2$ with
the help of the quadratic solution formula. The result is shown 
as the bottom solid line 
in Fig.~\ref{fig7}.
\end{enumerate}

%
\subsection{Doubly-approximated pseudopotential prediction}
\label{secIVc}

Explicit expressions of the boundaries $a_{\text{boundary}}(q)$
between rod and pop-out, and tilt and planar phases may be obtained
by further approximating $\tilde{U}_{\text{eff}}$ in (\ref{ueff3}) up to
first order in $a$ and employing the same methodology
used in Secs.~\ref{secIVa} and \ref{secIVb}.
Denoting the doubly-approximated improved potential as
$\tilde{U}^{(\text{app})}_{\text{eff}}$,
we obtain
\begin{align}
\label{uapp}
\tilde{U}^{(\text{app})}_{\text{eff}} = \left[ \frac{2}{\rho} + \frac{1}{2r} \right] 
+ \left[ \frac{3q^2}{2}+3a+\frac{3aq^2}{8} \right] x^2 + \left[ \frac{q^2}{2}+a+\frac{aq^2}{8} \right] y^2 
+ \left[ 2q^2-2a-aq^2 \right] z^2 \nonumber \\
+ \frac{q^2}{\rho^5} \bigg[ \frac{81x^4}{4}\left( 1+\frac{a}{2} \right) 
+ \frac{y^4}{4} \left( 1+\frac{a}{2} \right) + z^4 (1-a) 
+ \frac{9x^2y^2}{2} \left( 1+\frac{a}{2} \right)
-  \frac{y^2z^2}{8} \left( 17-\frac{13a}{2} \right) \nonumber \\
-  \frac{9z^2x^2}{8} \left( 17-\frac{13a}{2} \right) \bigg] 
+ \frac{q^2}{r^5} \left[ \frac{y^4}{16} \left(1+\frac{a}{2} \right)
+ \frac{z^4}{4}(1-a)-\frac{y^2z^2}{32} \left( 17-\frac{13a}{2} \right) \right].
\end{align}

(i) Rod $\rightarrow$ pop-out boundary

Repeating the above procedure, we obtain
\begin{equation}
\label{Vc1-1}
\frac{1}{z}\frac{\partial \tilde{U}^{(\text{app})}_{\text{eff}}(x=0,y=0,z)}{\partial z} = 
-\frac{5}{2z^3}+2(2q^2-2a-aq^2)-\frac{5(1-a)q^2}{4z^3}
\end{equation}
and
\begin{equation}
\label{Vc1-2}
\frac{\partial^2\tilde{U}_{\text{eff}}(x,y=0,z)}{\partial x^2} \Bigg|_{x=0} = 
2 \left(3 a + \frac{3 q^2}{2} + \frac{3 a q^2}{8}\right) - \frac{18}{z^3}
-\frac{9 \left(17 - \frac{13 a}{2}\right) q^2}{4 z^3} 
- \frac{45 (1 - a) q^2}{z^3}.
\end{equation}
Equating the two equations (\ref{Vc1-1}) and (\ref{Vc1-2}) to zeros
and eliminating $z$ results in
\begin{equation}
4 q^2 (86 + 439 q^2) + a^2 q^2 (1312 + 641 q^2) -
 a (464 + 2018 q^2 + 2145 q^4) = 0,
\end{equation}
which yields the small-$a$, $1/$length approximated
improved pseudopotential three-particle rod $\rightarrow$ pop-out boundary line
\begin{align}
a_{\text{rod}\rightarrow\text{pop-out}}(q) =&
\left[464 + 2018 q^2 + 2145 q^4 -
 (215296 + 1872704 q^2 + 4257572 q^4 \right. \nonumber \\
&\left. - 1440284 q^6 + 
  98641 q^8)^{1/2}\right]/(2624 q^2 + 1282 q^4).
\end{align}

(ii) Pop-out $\rightarrow$ tilt boundary

This time, we start with
\begin{align}
\label{Vc2-1}
\frac{1}{x}\frac{\partial \tilde{U}^{(\text{app})}_{\text{eff}}(x,y=0,z)}{\partial x} =& 
-\frac{18}{(9x^2+z^2)^{3/2}}+\left[ 3q^2+6a+\frac{3aq^2}{4} \right] \nonumber \\
&-\frac{45q^2}{(9x^2+z^2)^{7/2}}\left[ \frac{81x^4}{4}\left(1+\frac{a}{2}\right)
+z^4(1-a)-\frac{9x^2z^2}{8}\left(17-\frac{13a}{2}\right) \right] \nonumber \\
&+\frac{q^2}{(9x^2+z^2)^{5/2}}
\left[81x^2\left(1+\frac{a}{2}\right)-\frac{9z^2}{4}\left(17-\frac{13a}{2}\right)\right]\nonumber \\
=&0, \nonumber \\
\frac{1}{z}\frac{\partial \tilde{U}^{(\text{app})}_{\text{eff}}(x,y=0,z)}{\partial z} =& 
-\frac{1}{2z^3}-\frac{2}{(9x^2+z^2)^{3/2}}+\left[ 4q^2-4a-2aq^2 \right] \nonumber \\
&-\frac{5q^2}{(9x^2+z^2)^{7/2}}
\left[\frac{81x^4}{4}\left(1+\frac{a}{2}\right)+z^4(1-a)-\frac{9x^2z^2}{8}\left(17-\frac{13a}{2}\right)\right] 
\nonumber \\
&+\frac{q^2}{(9x^2+z^2)^{5/2}}\left[4z^2(1-a)-\frac{9x^2}{4}\left(17-\frac{13a}{2}\right)\right]
-\frac{q^2}{z^3}\frac{1-a}{4} \nonumber \\
=&0,
\end{align}
for the first derivatives, and
\begin{align}
\label{Vc2-3}
\frac{\partial^2 \tilde{U}^{(\text{app})}_{\text{eff}}(x,y,z)}{\partial y^2} \bigg|_{y=0} = 
&-\frac{1}{2z^3}-\frac{q^2}{16z^3}\left[\left(17-\frac{13a}{2}\right)+20(1-a)\right]
-\frac{2}{(9x^2+z^2)^{3/2}} \nonumber \\
&+\frac{q^2}{(9x^2+z^2)^{5/2}}\left[ 9\left(1+\frac{a}{2}\right)x^2
-\frac{1}{4}\left(17-\frac{13a}{2}\right)z^2\right] \nonumber \\
&-\frac{5q^2}{(9x^2+z^2)^{7/2}}\left[\frac{81}{4}\left(1+\frac{a}{2}\right)x^4
-\frac{9}{8}\left(17-\frac{13a}{2}\right)x^2z^2+(1-a)z^4\right] \nonumber \\
=&0
\end{align}
for the second derivative.
Subtracting $9$ times (\ref{Vc2-3}) from (\ref{Vc2-1}) results in
\begin{equation}
\label{Vc2-4}
z = \left[
\frac{3(16+74q^2-53aq^2)}
{16(8a+4q^2+aq^2)}
\right]^{1/3}.
\end{equation}
Subtracting $9$ times the second equation of (\ref{Vc2-1}) from
the first equation of (\ref{Vc2-1}) and rearranging terms
results in an equation of the form (\ref{Vb2-5}), where
\begin{align}
\label{Vc2-6}
\delta &= \mu_x^2 - 9\mu_z^2 + \frac{75}{4}aq^2+\frac{9}{2z^3}, \nonumber \\
\beta  &= -q^2\left[81\left(1+\frac{a}{2}\right) + \frac{81}{4}\left(17-\frac{13a}{2}\right)\right], \nonumber \\
\gamma &= q^2\left[\frac{9z^2}{4}\left(17-\frac{13a}{2}\right) +36z^2(1-a)\right].
\end{align}
As in the analogous case discussed in Sec.~\ref{secIVb},
following the detailed discussion in Appendix~\ref{appC},
we find 
\begin{equation}
x = z\sqrt{\frac{5}{9}-\frac{16}{42-9a}},
\end{equation}
which, together with $z$ in (\ref{Vc2-4}), results in
the pop-out $\rightarrow$ tilt boundary line
\begin{equation}
a_{\text{pop-out}\rightarrow\text{tilt}} = 
\frac{576 + 2288 q^2 + 2091 q^4 - \sqrt{
 331776 + 2635776 q^2 + 5934208 q^4 - 858592 q^6 + 
  58081 q^8}}{53 (56 q^2 + 25 q^4)}. 
\end{equation}

(iii) Planar $\rightarrow$ tilt boundary

For the third and last boundary, i.e., the one between
the tilt and planar phases, we begin with 
\begin{align}
\label{Vc3-1}
\frac{1}{x} \frac{\partial \tilde{U}^{(\text{app})}_{\text{eff}}}{\partial x} \Bigg|_{z=0,y=\sqrt{3}x} 
=&-\frac{\sqrt{3}}{2x^3} -\frac{\sqrt{3}q^2}{16x^3}\left(1+\frac{a}{2}\right) \nonumber \\
&+ 6\left(a+\frac{q^2}{2}+\frac{aq^2}{8}\right)+2\left(3a+\frac{3q^2}{2}+\frac{3aq^2}{8}\right) = 0
\end{align}
and
\begin{align}
\label{Vc3-2}
\frac{\partial^2 \tilde{U}^{(\text{app})}_{\text{eff}}}{\partial z^2}\Bigg|_{z=0,y=\sqrt{3}x}  
=& 2(-2a+2q^2-aq^2)-\frac{1}{4\sqrt{3}x^3} \nonumber \\
&-\frac{q^2}{32\sqrt{3}x^3}\left(17-\frac{13a}{2}\right)
-\frac{5q^2}{32\sqrt{3}x^3}\left(1+\frac{a}{2}\right)=0.
\end{align}
Eliminating $x$ from (\ref{Vc3-1}) and (\ref{Vc3-2}), we obtain
\begin{equation}
4q^2a^2+ 4 (4 - 3 q^2) q^2 - a [32 + q^2 (44 + q^2)]=0,
\end{equation}
which may be solved explicitly for $a$ to result in
\begin{equation}
a_{\text{tilt}\rightarrow\text{planar}} = \frac{32 + 44 q^2 + q^4 - \sqrt{
 1024 + 2816 q^2 + 1744 q^4 + 280 q^6 + q^8}}{8 q^2}.
\end{equation}

\begin{figure}
\includegraphics[scale=2.2,angle=0]{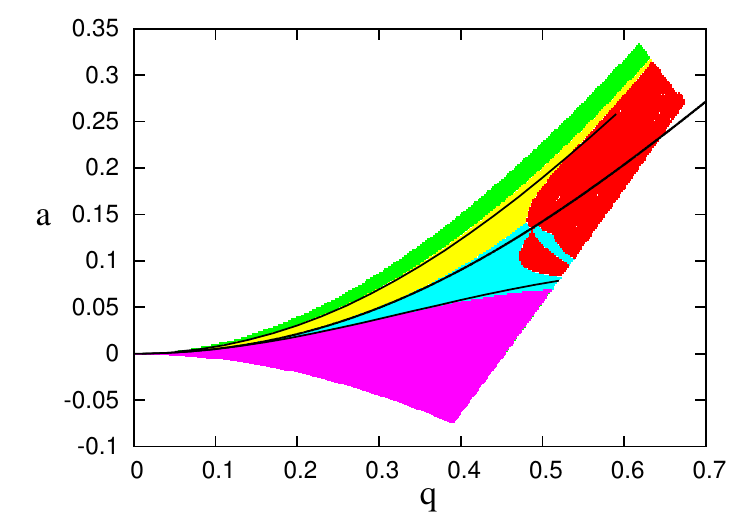}
\caption {\label{fig8}
(Color Online) 
Improved, 1/length, small-$a$ approximated
generalized three-ion pseudopotential prediction of morphology boundaries 
(solid lines) predicted in Sec.~\ref{secIVc} 
in the ($q,a$) control-parameter space.
Color codes for the three-particle morphologies
are the same as those used in Figs.~\ref{fig6} and \ref{fig7}.
}
\end{figure}

All boundaries (i)-(iii), together with our simulation results,
are presented in Fig.~\ref{fig8}.
The agreement between the predicted boundaries
and numerical results is excellent, in particular 
for small $a$.
For $q<0.4$, e.g., the analytical boundaries fall within
$5\%$ of the exact, numerically computed boundaries.

\section{Four-ion Crystal Morphologies}
\label{secV}

With increasing $N$, as is already apparent in Sec.~\ref{secIV} 
for the case of three particles,
analytical analysis of the crystal configurations becomes
an increasingly difficult task to perform.
Thus, in this section, where we focus on the case $N=4$, 
we primarily present numerical results.

\begin{figure}
\centering
\includegraphics[scale=0.2,angle=0]{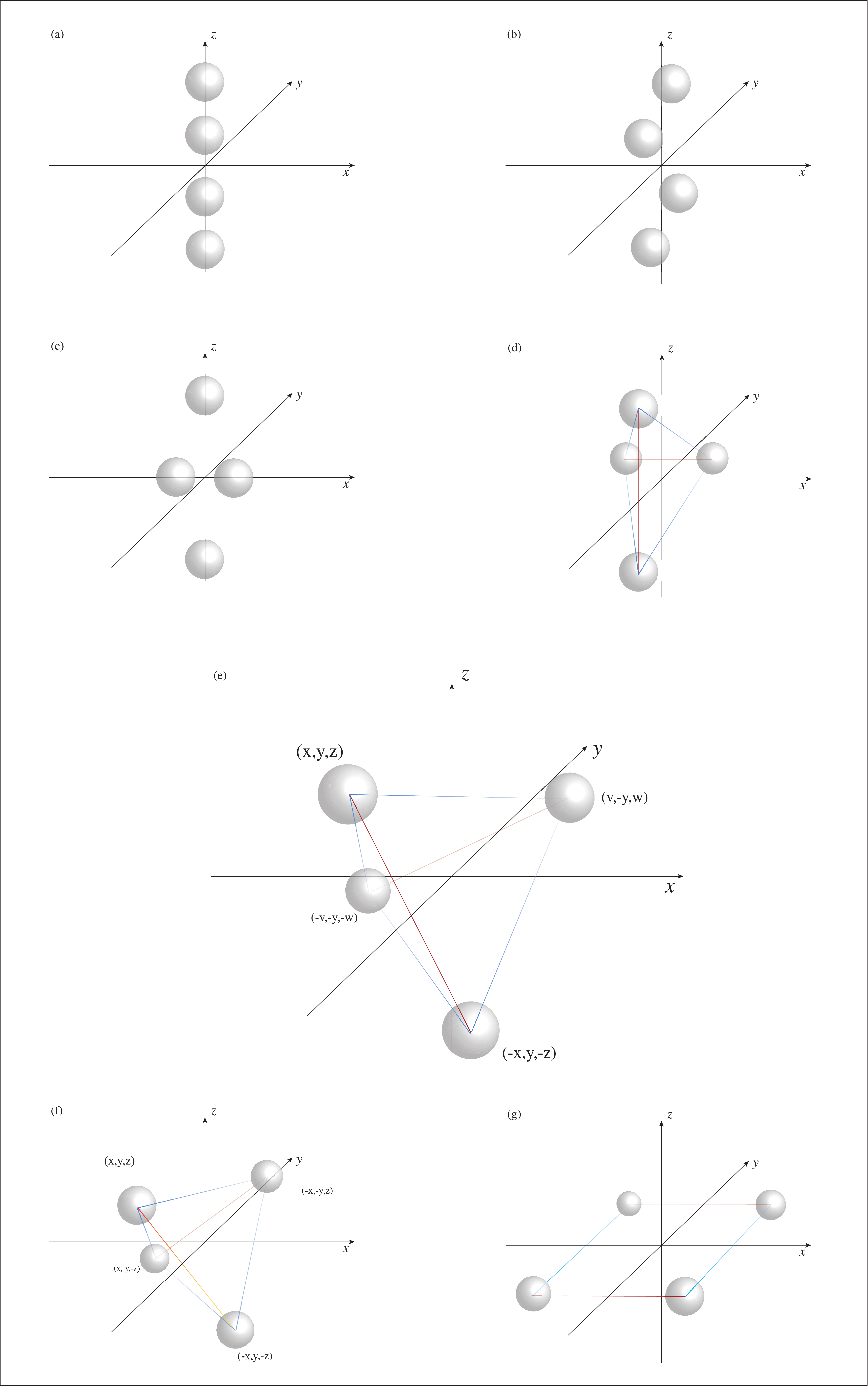}
\caption {\label{fig9}
The seven possible crystal configurations of four charged particles
stored in a Paul trap.
Panel~(e) represents the most general configuration,
wherein our position-parameter choice appears.
Lines drawn between particles in panels~(d)-(g)
are to guide the eye.
}
\end{figure}

Figure~\ref{fig9} shows the observed crystal orientations obtained from
our numerical simulations.
As detailed in Figs.~\ref{fig9}(a)-(g), 
we observe seven different morphologies. Among them, 
the most general configuration is shown in 
panel~(e),
which we magnified in relation to the other six 
morphologies illustrated in Fig.~\ref{fig9} to 
show our choice of position coordinates,
$x,y,z,v,$ and $w$, for the $N=4$ case.

\begin{figure}
\centering
\includegraphics[scale=1.5,angle=0]{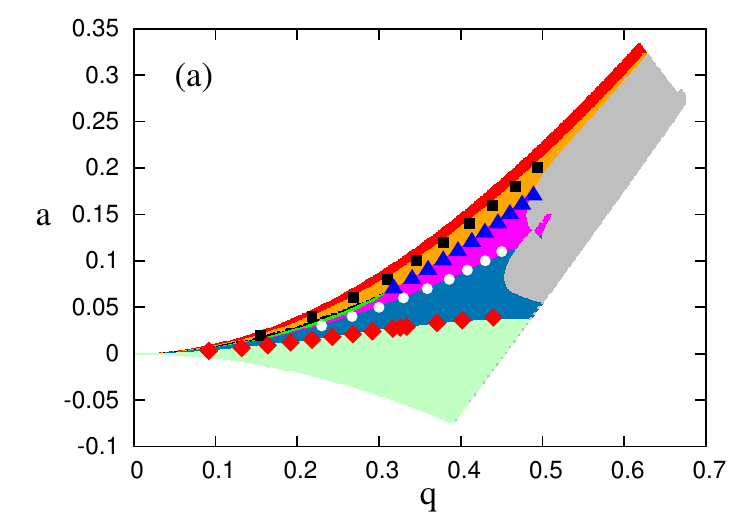} \\
\includegraphics[scale=1.5,angle=0]{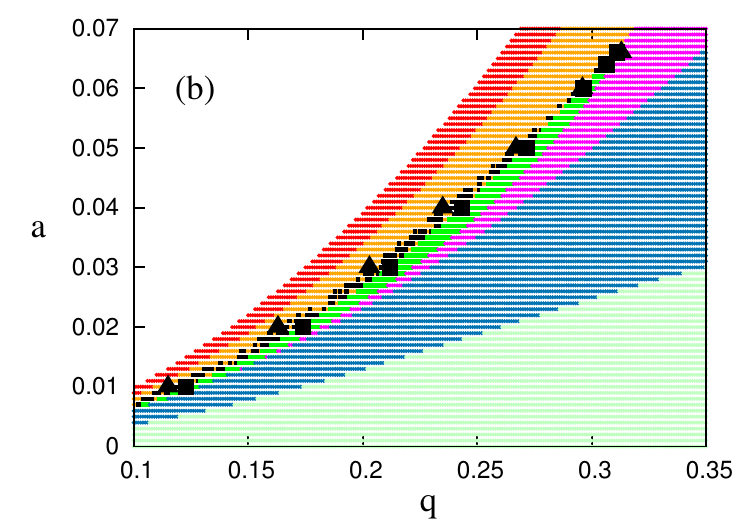}
\caption {\label{fig10} (Color online) 
Crystal morphologies according to our 
numerical simulations of  (\ref{TDEM3}) (colored regions) 
and their boundaries, predicted by the generalized
pseudopotential (plot symbols), for $N = 4$. 
The results for the full stability region are shown 
in (a); 
(b) shows a zoom of the region 
$0.1\leq q\leq 0.35$, $0\leq a\leq 0.07$.  
The morphologies shown in panels
(a)-(g) of Fig.~\ref{fig9} correspond to
the colors in the order of 
red, orange, black, green,
magenta, blue, and light green.
The analytical,   
generalized pseudopotential
predictions of the boundaries 
between the different morphologies
are indicated by the plot symbols in panels (a) and (b): 
squares, triangles, circles, and diamonds in (a)
indicate the red $\leftrightarrow$ orange,
orange $\leftrightarrow$ magenta,
magenta $\leftrightarrow$ blue,
and blue $\leftrightarrow$ light-green boundaries,
respectively. 
Triangles and squares in (b)
indicate orange $\leftrightarrow$ black
and black $\leftrightarrow$ green boundaries, respectively.
}
\end{figure}

Numerical results for locations and types of four-ion 
crystal morphologies 
are shown in Fig.~\ref{fig10} 
along with the numerical solutions of the 
boundaries between them, following the 
general analytical framework available in Appendix~\ref{appA},
applied to the four-particle case.
The stability assessment of each crystal configuration 
is based on 
the Hessian matrix \cite{NumRec,GR},  
defined as $H_{ij} = \partial^2 U_{\text{eff}} / \partial i \partial j$,
where $i,j \in \{x,y,z,v,w\}$.
Except for one, 
which corresponds to the metamorphosis
between the configurations shown in panels~(d) and (e) 
of Fig.~\ref{fig9}, all morphology boundaries were
determined via the Hessian method.
The numerically evaluated analytical boundaries
match the results of our 
numerical simulations of (\ref{TDEM3}) to an excellent degree 
of accuracy. 

\begin{figure}
\centering
\includegraphics[scale=1.6,angle=0]{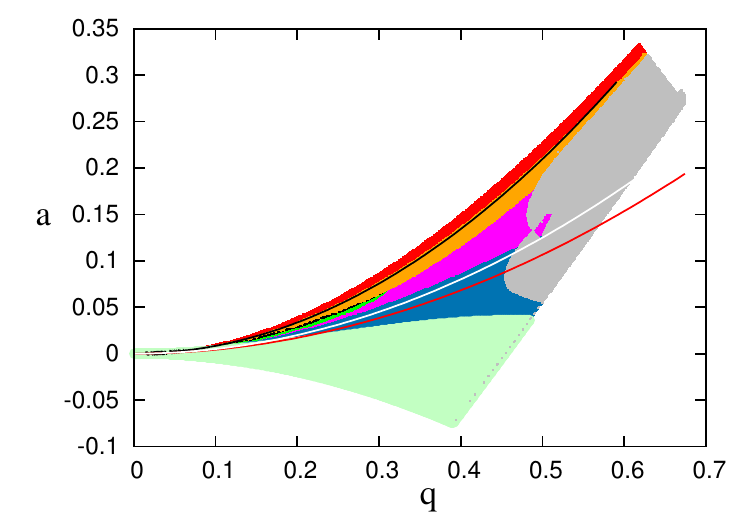}
\caption {\label{fig11} (Color online) 
$N = 4$
crystal morphologies (colored regions) 
and their boundaries (solid lines) 
predicted by the standard
pseudopotential.
The color codes for the morphologies 
are identical to those in Fig.~\ref{fig10}.
The black line denotes the prediction of 
the standard pseudopotential for the 
boundary between red and orange;
the white line, $a = q^2/2$, denotes
the (degenerate) prediction of the standard 
pseudopotential for the 
boundaries between orange and black,
orange and magenta, and magenta and blue;
the red line denotes the predicted 
boundary between blue and light green.
The predicted boundary between black and green
lies outside the ($q,a$) control-parameter 
region shown in the figure.
}
\end{figure}

In contrast, in Fig.~\ref{fig11}, we show the 
boundary lines expected from the standard pseudopotential framework.
As in the case of the generalized pseudopotential, we used the 
Hessian to numerically evaluate the resulting analytical boundaries.
Once again, our method results in all morphology boundaries
except for the one that corresponds to the metamorphosis between 
the configurations shown in panels~(d) and (e) of Fig.~\ref{fig9}.
Expectedly so, and in contrast to our improved pseudopotential, 
the standard pseudopotential fails to predict
many of the boundaries observed in our numerical simulations.
Once again, this corroborates the power of the generalized pseudopotential,
accurately predicting the morphology boundaries 
of four-ion crystals in the Paul trap.

\section{Discussion}
\label{secVI}
At the beginning of the 20th century, prompted by the 
eminent mathematician David Hilbert (of Hilbert-space fame), 
F\"oppl \cite{Foeppl} was the first to investigate analytically  
crystalline structures of charged particles in a harmonic-oscillator
potential. At that time, the physical 
application of charged-particle crystals was not to traps 
but to atomic structure, since, according to 
Thomson's plum-pudding model, the prevailing atomic model 
at the beginning of the 20th century, 
electrons were thought to be embedded 
in a positively charged sphere, and the groundstate of atoms was 
thought to be formed by the crystalline structures of electrons 
in this positive background charge. 
Since, according to Maxwell's theory, already highly developed 
at the time, a homogeneously charged sphere corresponds to 
a harmonic oscillator potential, by computing 
crystalline electron structures in an isotropic oscillator 
potential, F\"oppl had, in fact, solved the 
atomic structure problem of his time. 
While, these days, our atomic models differ quite a bit from 
the atomic models prevailing at F\"oppl's time, in connection 
with quantum dots \cite{QDOT1}, sometimes called 
``artificial atoms'' \cite{QDOT2}, 
Thomson's plum-pudding model has recently experienced a renaissance. 
Quantum dots have numerous applications in physics and 
technology since, among others applications, 
they may act as transistors, solar cells, and lasers 
\cite{QDOT1}.
In essence, a 
quantum dot is a nano-scale semiconductor trap, capable of 
confining electrons in all three spatial dimensions. 
To lowest order, the electric potential confining 
the electrons in the quantum 
dot is a harmonic oscillator potential. Thus, a 
quantum dot may be considered a realization 
of Thomson's plum-pudding model, which, unexpectedly, 
returns F\"oppl's work \cite{Foeppl}, published more than 
a century ago, to the front burner. 
 
F\"oppl's electron crystals are relevant in our context, 
since they describe the crystalline structures expected 
in what we call the standard pseudopotential. Our work, 
however, reported in this paper, goes beyond 
F\"oppl's work in that we consider crystalline structures 
in the cylindrically symmetric, but highly deformed,  
improved pseudopotentials. For these potentials, 
we succeeded to compute and classify the 
structures of all two-, three-, and 
four-ion crystals and their boundaries as a function 
of the Paul-trap control parameters $a$ and $q$. 
 
Unlike F\"oppl's work \cite{Foeppl}, or the work on 
quantum dots \cite{QDOT1,QDOT2}, our starting point is 
not a time-independent potential. Quite the contrary. 
Our starting point are the periodically driven, 
time-dependent equations of 
motion (\ref{TDEM3}), for which a time-independent 
potential first has to be derived. In the Introduction 
of this paper (Sec.~\ref{INTRO}) we saw that, 
for periodically driven quantum systems, 
in the stroboscopic picture, a time-independent 
Hamiltonian always exists (the quasi-energy operator), 
but that for nonlinear, classical systems, the existence 
of a meaningful time-independent Hamiltonian, in analogy 
to the quantum quasi-energy operator, cannot be 
guaranteed. Thus, by constructing the higher-order 
improved pseudopotentials and by showing that 
the crystals structures obtained in the pseudopotential 
picture reproduce all crystal structures seen in the 
(numerically) exact simulations of (\ref{TDEM3}), we 
showed that pseudopotentials for periodically driven, nonlinear 
systems can indeed be constructed in a systematic way. 

Loosely speaking, we carried our systematic expansions 
``up to second order'' by including not only $\cos(2\tau)$ 
terms in our analytical derivations, which yields the 
standard pseudopotential, but also $\cos(4\tau)$ terms, which, 
upon averaging, yield our improved pseudopotentials. It is 
this improvement step that is capable of capturing 
all crystal configurations, in contrast with the 
standard, lower-order standard pseudopotential, which did not 
catch many of the more ``exotic'' crystal morphologies. 
In addition, the boundary lines between different 
crystal morphologies are near perfectly predicted 
by our improved pseudopotential. 
 
However, there is one essential limitation of our 
improved pseudopotential. Even our improved pseudopotential 
does not predict 
the existence of the red regions in Figs.~\ref{fig4}, 
\ref{fig6}, \ref{fig7}, and \ref{fig8}, 
where no crystals exist. This may be beyond the power 
of even higher-order pseudopotentials, since the 
red regions in Figs.~\ref{fig4}, 
\ref{fig6}, \ref{fig7}, and \ref{fig8} represent 
regions of dynamical chaos \cite{LiLi} which are hard to 
represent with smooth potentials. This is not to say 
that time-independent potentials cannot produce chaos. 
Quite the contrary. Many examples are known, for 
instance the potential of the diamagnetic hydrogen atom 
\cite{HFr}
in AMO physics, which produces chaotic motions of the 
hydrogenic electron in a combination of the 
Coulomb field and an externally applied, static magnetic field. 
However, our potentials are different. They are supposed 
to accurately represent the macromotion positions at 
the stroboscopic times of the system, a task that 
is impossible to perform in the case of chaotic motion unless the potential itself 
has a ``chaotic'' structure. Nevertheless, one 
important property of the 
red regions in Figs.~\ref{fig4}, 
\ref{fig6}, \ref{fig7}, and \ref{fig8} is that 
as a prerequisite for the absence of crystals in 
these regions, period-one fixed points \cite{LiLi} 
need to 
lose their stability. This should be within the 
power of the pseudopotential approach, but needs to 
await expansions to even higher orders, since explicit 
investigations of fixed-point stability in the red 
regions show that even our improved pseudopotential 
still predicts (erroneously) stability of period-one 
fixed points in this region. A 
period-one fixed-point analysis on the basis of 
the full time-dependent equations of motion (\ref{TDEM3}) 
showed explicitly that these fixed points are 
indeed unstable in the 
red regions in Figs.~\ref{fig4}, 
\ref{fig6}, \ref{fig7}, and \ref{fig8}. 
 
%
%

\section{Summary and Conclusion}
\label{secVII}
One can be seriously misled by the predictions of 
the standard Paul-trap pseudopotential, which, e.g., fails to 
predict the existence of any of the more ``exotic'' 
ion crystal configurations in the Paul trap. This is 
the primary reason for why, in this paper, we developed 
an improved pseudopotential, whose predictive power was 
tested for two, three, and four simultaneously stored ions 
in the Paul trap. Unlike the standard pseudopotential, 
used by many researchers, the improved pseudopotential 
is capable of predicting, without fail, any of the many 
three- and four-ion crystal morphologies in the Paul trap, 
including the boundaries between these crystal structures 
in the $(q,a)$ control-parameter space of the Paul trap. 
Most of the three- and four-particle crystal structures predicted in our paper have 
never been seen experimentally. Many laboratories throughout 
the world have the ability to investigate and image few-ion 
systems in the Paul trap. We are looking forward to the  
experimental verification of the ion-crystal morphologies 
and the transitions between them 
predicted in this paper.

\appendix

\section{$n$-coordinate framework}
\label{appA}

In this appendix, we derive a general, $n$-coordinate
analytical framework that may be used to result in 
the improved pseudopotential expressions for any number
of particles in the Paul trap.
We start with the $n$ coupled differential equations
\begin{equation}
\label{A1}
m_{X^{(i)}}\ddot{X}^{(i)} = -U_{X^{(i)}} (X^{(1)},X^{(2)}, \ldots, X^{(n)}) 
					- k_{X^{(i)}}X^{(i)}\cos(\omega t),
\end{equation}
where $m_{X^{(i)}}$ are the effective masses corresponding to the coordinates $X^{(i)}$,
$U_{X^{(i)}} = \partial U/ \partial X^{(i)}$, and $k_{X^{(i)}}$ are constants.
Denoting
\begin{equation}
\label{A2}
X^{(i)}(t) = x^{(i)}(t) + \xi^{(i)}\cos(\omega t),
\end{equation}
assuming $\xi^{(i)}$ is small compared to $x^{(i)}(t)$,
and $x^{(i)}(t)$ varies slowly compared to the timescale of $\cos(\omega t)$,
we may expand $U_{X^{(i)}} (X^{(1)},X^{(2)},\ldots, X^{(n)})$ in $\xi^{(i)}(t)$
according to
\begin{align}
\label{A3}
U(X^{(1)},X^{(2)},\ldots, X^{(n)}) 
&\approx U(x^{(1)},x^{(2)},\ldots, x^{(n)}) \nonumber \\
&+\sum_{i=1}^n U_{x^{(i)}}\xi^{(i)}\cos(\omega t)
+\sum_{i,j}\frac{1}{2} U_{x^{(i)}x^{(j)}}\xi^{(i)}\xi^{(j)} \cos^2(\omega t),
\end{align}
where $U_{x^{(i)}x^{(j)}} = \partial^2 U / \partial x^{(i)} \partial x^{(j)}$,
evaluated at $X^{(i)} = x^{(i)}$ and $X^{(j)} = x^{(j)}$.
With (\ref{A3}), we may write
\begin{align}
\label{A4}
U_{X^{(i)}} &= U_{x^{(i)}} + \sum_{j=1}^{n} U_{x^{(j)}}x^{(i)} \xi^{(j)} \cos(\omega t) \nonumber \\
&+\frac{1}{2}\sum_{j,k} U_{x^{(k)}x^{(j)}x^{(i)}}\xi^{(k)}\xi^{(j)} \cos^2 (\omega t),
\end{align}
which may then be inserted into (\ref{A1}) to result in 
\begin{equation}
\label{A5}
m_{X^{(i)}}\omega^2\xi^{(i)} = \sum_{j=1}^n U_{x^{(j)}x^{(i)}}\xi^{(j)} + k_{X^{(i)}}x^{(i)}
\end{equation}
upon comparing coefficients of corresponding powers of 
$\cos(\omega t)$.
Since (\ref{A5}) is a system of $n$ coupled linear equations in $\xi^{(i)}$,
denoting $\mathcal{A}$ as the matrix whose matrix elements are
\begin{align}
\label{A6}
\mathcal{A}_{ij} = 
\begin{cases}
m_{X^{(i)}}\omega^2-U_{x^{(i)}x^{(i)}} &\mbox{if   } i=j, \\
-U_{x^{(j)}x^{(i)}} &\mbox{if   } i\neq j,
\end{cases}
\end{align}
and denoting $\vec{\xi} = (\xi^{(1)},\xi^{(2)},\ldots,\xi^{(n)})^{\text{T}}$
and $\vec{b} = (k_{X^{(1)}}x^{(1)},k_{X^{(2)}}x^{(2)},\ldots,k_{X^{(n)}}x^{(n)})^{\text{T}}$,
we obtain the $i$th element $\xi^{(i)}$ of $\vec{\xi}$ 
using Cramer's rule \cite{Cramer}   
\begin{equation}
\label{A7}
\xi^{(i)} = \frac{\text{det}(\mathcal{A}_i)}{\text{det}(\mathcal{A})},
\end{equation}
where $\text{det}(...)$ denotes the determinant of the matrix argument
and $\mathcal{A}_i$ is the matrix whose $i$th column is $\vec{b}$
but otherwise the same as matrix $\mathcal{A}$.
Averaging (\ref{A1}) over one cycle of $\cos(\omega t)$,
with (\ref{A7}), the effective equations of motion become
\begin{equation}
m_{X^{(i)}}\ddot{x}^{(i)} = -U_{x^{(i)}} - \frac{1}{4} \sum_{j,k} U_{x^{(k)}x^{(j)}x^{(i)}} \xi^{(k)}\xi^{(j)} 
-\frac{1}{2}k_{X^{(i)}}\xi^{(i)},
\end{equation}
which can be shown by direct differentiation 
to have the effective potential 
\begin{equation}
U_{\text{eff}}(x^{(1)},x^{(2)},\ldots,x^{(n)}) = U(x^{(1)},x^{(2)},\ldots,x^{(n)})
+\frac{1}{4}\sum_{i} k_{X^{(i)}}x^{(i)}\xi^{(i)}.
\end{equation}

\section{Inverse-length approximation of the pseudopotential}
\label{appB}

In this appendix we present the detailed steps to arrive at
the inverse-length approximation of the 
pseudopotential $\tilde{U}_{\text{eff}}$
in (\ref{ueff3}), starting from the pseudopotential $U_{\text{eff}}$ in (\ref{ueff2}).
For the convenience of the reader, we import $U_{\text{eff}}$ in (\ref{ueff2}):
\begin{equation}
\label{B1}
U_{\text{eff}} = \frac{2}{\rho}+\frac{1}{2r}+3ax^2+ay^2-2az^2
                          +\frac{3q x\xi\Delta + q y\eta\Delta - 2q z\zeta\Delta}{\Delta}.
\end{equation}

Since the goal is to keep terms up to first order in $1/\text{length}$,
all we need to do is to expand the last term in (\ref{B1}).
Denoting the numerator of this term by $N$ 
and the denominator of this term by $D$,
we see that, together with (\ref{micro}), (\ref{microD}), and (\ref{partial-table}),
$N \in O(\text{length}^2)$ and $D \in O(1)$, as $\text{length} \rightarrow +\infty$.
This means, (i) since $D$ is of the order of a constant in length,
$N$ should be expanded up to $1/\text{length}$ and
(ii) since $N$ is of the order of $\text{length}^2$, 
$D$ must be expanded up to $1/\text{length}^3$.
This is so, because
\begin{align}
\label{B2}
\frac{N}{D} &= \frac{a_2+a_1+a_0+a_{-1} +O(\text{length}^{-2})}
                                  {b_0 +b_{-1} +b_{-2} +b_{-3} +O(\text{length}^{-4})} \nonumber \\
        &\approx \frac{1}{b_0}\left[a_2+a_1+a_0+a_{-1}+O(\text{length}^{-2})\right] \nonumber \\
             &\times \left\{1-\frac{1}{b_0}\left[b_{-1}+b_{-2}+b_{-3}+O(\text{length}^{-4})\right]\right\}^{-1},
\end{align}
where $a_\nu$ and $b_\mu$ denote the terms that scale like
$\text{length}^\nu$ and $\text{length}^\mu$ of $N$ and $D$, respectively.

Direct evaluation of $a_\nu$ and $b_\mu$ 
for $\nu = 2,1,0,-1$ and $\mu=0,-1,-2,-3$ results in 
\begin{align}
\label{B3}
a_1 =& a_0 = b_{-1} = b_{-2} = 0, \nonumber \\ 
a_2 =& 36q^2x^2(8-2a)(8+4a) + 4q^2y^2(8+4a)(24-6a)+16q^2z^2(24-6a)(8-2a), \nonumber \\ 
b_0 =& (8-2a)(8+4a)(24-6a), \nonumber \\ 
a_{-1} = 
           &3qx\left\{12qx\left[
              (8 - 2 a)\left(\frac{1}{2 r^3} +\frac{2}{\rho^3} 
              - \frac{6 z^2}{\rho^5}-\frac{3 z^2}{2 r^5}\right) \right.\right. \nonumber \\
           &\qquad\qquad\;\left.+(8 + 4 a)\left(\frac{1}{2 r^3} + \frac{2}{\rho^3} 
              - \frac{6 y^2}{\rho^5} - \frac{3 y^2}{2 r^5}\right)\right] \nonumber \\
           &\quad\:\:\:\,+ 4qy\left[(8 + 4 a)\left(\frac{54xy}{\rho^5}\right)\right]
              \left.-8qz\left[(8 - 2 a)\left(\frac{54xz}{\rho^5}\right)\right]\right\} \nonumber \\
         +&qy\left\{4qy\left[(8 + 4 a)\left(\frac{18}{\rho^3} - \frac{486 x^2}{\rho^5}\right) 
           + (24 - 6 a)\left(\frac{1}{2 r^3} + \frac{2}{\rho^3} 
           - \frac{6 z^2}{\rho^5} - \frac{3 z^2}{2 r^5}\right)\right]\right. \nonumber \\
           &\quad\:\:\left.+12qx\left[(8 + 4 a)\left(\frac{54xy}{\rho^5}\right)\right]
           -8qz\left[(24 - 6 a)\left(\frac{6yz}{\rho^5} + \frac{3yz}{2 r^5}\right)\right]\right\} \nonumber \\
           -&2qz\left\{-8qz\left[(24 - 6 a)\left(\frac{1}{2 r^3} + \frac{2}{\rho^3}
            -\frac{6 y^2}{\rho^5} - \frac{3 y^2}{2 r^5}\right) 
            +(8-2a)\left(\frac{18}{\rho^3}-\frac{486 x^2}{\rho^5}\right)\right]\right. \nonumber \\
           &\quad\:\:\left.+12qx\left[(8 - 2 a)\left(\frac{54xz}{\rho^5}\right)\right]
           + 4qy\left[(24 - 6 a)\left(\frac{6yz}{\rho^5} + \frac{3yz}{2 r^5}\right)\right]\right\},\nonumber \\
b_{-3} =&  (8-2a)(8+4a)\left(-\frac{486x^2}{\rho^5}+\frac{18}{\rho^3}\right) \nonumber \\
              &+(8-2a)(24-6a)\left(-\frac{3z^2}{2r^5}+\frac{1}{2r^3}
                                                   -\frac{6z^2}{\rho^5}+\frac{2}{\rho^3}\right) \nonumber \\
              &+(8+4a)(24-6a)\left(-\frac{3y^2}{2r^5}+\frac{1}{2r^3}
                                                    -\frac{6y^2}{\rho^5}+\frac{2}{\rho^3}\right). 
\end{align}
Keeping terms up to $1/\text{length}$ in (\ref{B2}),
together with the results shown in (\ref{B3}), we obtain
\begin{equation}
\frac{N}{D} \approx \frac{a_2+a_{-1}}{b_0} - \frac{a_2 b_{-3}}{b_0^2},
\end{equation}
which can be inserted in place of the last term in (\ref{B1})
to result in the improved pseudopotential $\tilde{U}_{\text{eff}}$ in (\ref{ueff3}).

\section{Analytical prediction of the 
thee-particle pop-out$\leftrightarrow$tilt boundary}
\label{appC}

In this appendix we fill in the steps in the calculation of 
the explicit solution of $x$ for determining 
the pop-out$\leftrightarrow$tilt boundary.
We start with (\ref{Vb2-5}), 
where, for fixed $a$ and $q$, 
the coefficients $\delta$, $\beta$, $\gamma$, 
and $z$ are constants. 
Introducing the variable $\psi$ to denote $x^2$, 
and squaring both sides of (\ref{Vb2-5}), 
we obtain
\begin{equation}
\delta^2(9\psi+z^2)^5 = (\beta\psi+\gamma)^2,
\end{equation}
which is a quintic equation in $\psi$.
A general formula that yields the solutions of the quintic 
equation does not exist \cite{quint-gen}.  
However, we note that, together with the transformation 
\begin{equation}
9\psi+z^2 \rightarrow \varphi,
\end{equation}
we may write
\begin{equation}
\delta^2\varphi^5 - \frac{\beta^2\varphi^2}{81}
+\frac{2}{9}\left(\frac{\beta z^2}{9}-\gamma\right)\beta\varphi 
- \left(\frac{\beta z^2}{9}-\gamma\right)^2 = 0,
\end{equation}
which is known as the principal quintic form.
By rescaling $\varphi$ according to
\begin{equation}
\label{C4}
\bar\varphi = -\frac{2\beta}{9}\frac{1}{\beta z^2/9-\gamma}\varphi , 
\end{equation}
we obtain
\begin{equation}
\label{C5}
A{\bar\varphi}^5+B{\bar\varphi}^2+\bar\varphi+1 = 0,
\end{equation}
where
\begin{equation}
\label{C6}
A = \left(\frac{9}{2\beta}\right)^5\delta^2\left(\frac{\beta z^2}{9}-\gamma\right)^3, \quad
B = \frac{1}{4}.
\end{equation}

Now, together with $z$ in (\ref{Vb2-4}) and $\delta$, $\beta$, and $\gamma$ in (\ref{Vb2-5}), 
we may numerically evaluate the coefficient $A$ in (\ref{C6}) explicitly as a function of the trap control parameters
$a$ and $q$. We find that, throughout the entire stability region, $|A| \sim 10^{-8} \ll 1$. 
This means that we may approximate the quintic term in (\ref{C5}) as 0 and solve the resulting quadratic equation, 
an excellent approximation if the solution $\bar\varphi_0$ of the quadratic equation is such that $|A|\bar\varphi_0 \ll 1$. 
It can straightforwardly be shown that $\bar\varphi_0 = -2$, and thus the approximation is indeed excellent. As a result, we obtain
\begin{equation}
\label{C7}
x = \sqrt{-\frac{\gamma}{\beta}}.
\end{equation}


\begin{thebibliography}{99}

\bibitem{AE} L. Allen and J. H. Eberly, 
{\it Optical Resonance and Two-Level Atoms} 
(Dover Publications, New York, 1987). 
 
\bibitem{SSL} M. Sargent, M. O. Scully, W. E. Lamb, 
{\it Laser Physics} (Addison-Wesley, Reading, MA, 1974). 
 
\bibitem{BK} J. E. Bayfield and P. M. Koch, 
Phys. Rev. Lett. {\bf 33}, 258 (1974); 
 
\bibitem{KL} P. M. Koch and K. A. H.van Leeuwen, 
Phys. Rep. {\bf 255}, 289 (1995).

\bibitem{RB1} 
R. Bl\"umel and U. Smilansky, 
Physica Scripta {\bf 35}, 15
     (1987).
 
\bibitem{RB2} R. Bl\"umel, A. Buchleitner, R. Graham, L. Sirko, 
U. Smilansky, and H. Walther, 
Phys. Rev. A {\bf 44}, 4521
    (1991).
    
\bibitem{BFS} R. Bl\"umel, S. Fishman, and U. Smilansky, 
J. Chem. Phys. {\bf 84}, 2604
      (1986). 
 
\bibitem{Paul1} W. Paul, 
Rev. Mod. Phys. {\bf 62}, 531
(1990). 
 
\bibitem{PKG} P. K. Ghosh, {\it Ion Traps} 
(Clarendon Press, Oxford, 1995). 
 
\bibitem{Nature}   R. Bl\"umel, J. M. Chen, E. Peik, W. Quint, 
W. Schleich, Y. R. Shen, 
and H. Walther, 
Nature {\bf 334}, 309
(1988).  
 
\bibitem{QE}  A. G. Fainshtein, N. L. Manakov, and L. P. Rapoport, 
J. Phys. B: Atom. Molec. Phys. {\bf 11}, 2561 (1978).  

\bibitem{EMP} {\it Atomic, Molecular, and Optical Physics Handbook}, 
edited by G. W. F. Drake (American Institute of Physics, 
Woodbury, New Yrok, 1996). 
 
\bibitem{Floquet} G. Floquet,
Annales de l'ƒcole Normale SupŽrieure
{\bf 12}, 47 (1883).
 
\bibitem{AS} {\it Handbook of Mathematical Functions},
edited by M. Abramowitz and I. A. Stegun
(National Bureau of Standards, Gaithersburg, MD, 1964).

\bibitem{Stock}
H.-J. St\"{o}ckmann, 
{\it Quantum Chaos: An Introduction} 
(Cambridge University Press, New York,1999).
 
\bibitem{QCR} R. Bl\"umel and W. P. Reinhardt, 
{\it Chaos in Atomic Physics} 
(Cambridge University Press, Cambridge, 1997).
 
\bibitem{AMOQE}  X. Chu and Shih-I Chu,
Phys. Rev. A {\bf 63}, 013414 (2000).

 
\bibitem{SPECP} M. Combescure,
J. Stat. Phys. {\bf 59}, 679 (1990).

\bibitem{LL} L. D. Landau and E. M. Lifshits, {\it Mechanics}, 
second edition (Pergamon Press, Oxford, 1976). 
 
\bibitem{2ION} M. G. Moore and R. Bl\"umel,
Phys. Rev. A {\bf 50}, R4453
(1994).
 
\bibitem{HB1} J. A. Hoffnagle and R. G. Brewer, 
Appl. Phys. B {\bf 60}, 113 
(1995). 

\bibitem{MB} M. G. Moore and R. Bl\"umel,
Physica Scripta T{\bf 59}, 429
(1995).

\bibitem{BlWe} R. Blatt and G. Werth,
Phys. Rev. A {\bf 25}, 1476 
(1982).
                    
\bibitem{WSL} R. F. Wuerker, H. Shelton, and R. V. Langmuir, 
J. Appl. Phys. {\bf 30}, 342
(1959). 

\bibitem{MPQ1} F. Diedrich, E. Peik, J. M. Chen, W. Quint, and H. Walther, 
Phys. Rev. Lett. {\bf 59}, 2931
(1987). 
 
\bibitem{NIST1} D. J. Wineland, J. C. Bergquist, W. M. Itano, 
J. J. Bollinger, and C. H. Manney, 
Phys. Rev. Lett. {\bf 59}, 2935
(1987). 
 
\bibitem{Wineland1} 
F. Diedrich, J. C. Bergquist, W. M. Itano, and D. J. Wineland, 
Phys. Rev. Lett. {\bf 62}, 403 (1989). 
 
\bibitem{EBJ} Y. S. Nam, E. B. Jones, and R. Bl\"umel,
Phys. Rev. A {\bf 90}, 013402 (2014).
 
\bibitem{Kappler} R. Bl\"umel, C. Kappler, W. Quint, and H. Walther, 
Phys. Rev. A {\bf 40}, 808
(1989). 
 
\bibitem{NumRec} W. H. Press, S. A. Teukolsky, 
W. T. Vetterling, and B. P. Flannery, 
{\it Numerical Recipes}, second edition 
(Cambridge University Press, Cambridge, 1992). 
 
\bibitem{Walth} H. Walther, 
    Advances in Atomic, Molecular, and Optical Physics 
    {\bf 31}, 137 
    (1993).
    
\bibitem{dim} J. D. Tarnas, Y. S. Nam, and R. Bl\"umel,
Phys. Rev. A {\bf 88}, 041401 (2013).

\bibitem{Euler} S. T. Thornton and J. B. Marion,
{\it Classical Dynamics of Particles and Systems},
(Brooks/Cole, Belmont, CA, 2004).
 
\bibitem{EMB} J. W. Emmert, M. Moore, and R. Bl\"umel, 
    Phys. Rev. A {\bf 48}, R1757
    (1993).
    
\bibitem{GR} I. S. Gradshteyn and I. M. Ryzhik, 
{\it Table of Integrals, Series, and Products}, 
fifth edition, A. Jeffrey, editor (Academic Press, 
Boston, 1994), \P 14.314, p. 1145. 
 
\bibitem{Foeppl} L. F\"oppl, 
J. reine angew. Math. {\bf 141}, 251
(1912). 
 
\bibitem{QDOT1} {\it Quantum Dots - Optics, Electron 
Transport and Future Applications}, 
edited by A. Tartakovskii 
(Cambridge University Press, Cambridge, 2012). 
 
\bibitem{QDOT2} M. A. Kastner, 
Physics Today {\bf 46} (1), 24
(1993). 
 
\bibitem{LiLi} A. J. Lichtenberg and M. A. Lieberman, 
{\it Regular and Stochastic Motion}, Vol. {\bf 38} 
of Applied Mathematical Sciences (Springer, 
New York, 1983). 
 
\bibitem{HFr} H. Friedrich, 
{\it Theoretical Atomic Physics}, third edition 
(Springer, Berlin, 2006). 
 
\bibitem{Cramer} D. C. Lay,
{\it Linear Algebra and its Applications}
(Addison-Wesley, New York, 2003).
  
\bibitem{quint-gen} N. Jacobson,
{\it Basic Algebra I} (Dover, Mineola, NY, 2009).
 
\end{thebibliography}
\end{document}